\documentclass[journal=jacsat,manuscript=article]{achemso}

\usepackage{chemformula} % Formula subscripts using \ch{}
\usepackage[T1]{fontenc} % Use modern font encodings
\usepackage{array}

\author{Melissa Bosch}
\affiliation{Department of Physics, Cornell University, Ithaca, New York 14853, USA}

\author{Maxim R. Shcherbakov}
\affiliation{School of Applied and Engineering Physics, Cornell University, Ithaca, New York 14853, USA}
\email{mrs356@cornell.edu}

\author{Kanghee Won}
\affiliation{Samsung Advanced Institute of Technology (SAIT), Samsung Electronics, Co. Ltd., Suwon-si, Gyeonggi-do 16678, Republic of Korea}

\author{Hong-Seok Lee}
\affiliation{Samsung Advanced Institute of Technology (SAIT), Samsung Electronics, Co. Ltd., Suwon-si, Gyeonggi-do 16678, Republic of Korea}

\author{Young Kim}
\affiliation{Samsung Advanced Institute of Technology (SAIT), Samsung Electronics, Co. Ltd., Suwon-si, Gyeonggi-do 16678, Republic of Korea}

\author{Gennady Shvets}
\affiliation{School of Applied and Engineering Physics, Cornell University, Ithaca, New York 14853, USA}
\title{Electrically actuated varifocal lens based on liquid-crystal-embedded dielectric metasurfaces}

\keywords{Metalens, liquid crystals, tunable optics, design and optimization}

\begin{document}

\maketitle

\begin{abstract}
Compact varifocal lenses are essential to various imaging and vision technologies. However, existing varifocal elements typically rely on mechanically-actuated systems with limited tuning speeds and scalability. Here, an ultrathin electrically controlled varifocal lens based on a liquid crystal (LC) encapsulated semiconductor metasurface is demonstrated. Enabled by the field-dependent LC anisotropy, applying a voltage bias across the LC cell modifies the local phase response of the silicon meta-atoms, in turn modifying the focal length of the metalens. In a numerical implementation, a voltage-actuated metalens with continuous zoom and up to 20$\%$ total focal shift is demonstrated. The concept of LC-based metalens is experimentally verified through the design and fabrication of a bifocal metalens that facilitates high-contrast switching between two discrete focal lengths upon application of a 3.2~V$_{\rm pp}$ voltage bias. Owing to their ultrathin thickness and adaptable design, LC-driven semiconductor metasurfaces open new opportunities for compact varifocal lensing in a diversity of modern imaging applications.
\end{abstract}

\section{Introduction}

Many optical systems rely on varifocal optical elements, which enable focusing at different planes of an imaged scene. Traditionally, optical zoom systems have consisted of moving lenses, an approach that requires large-footprint, mechanically complex solutions, and it remains an open challenge to realize a miniature non-mechanically-actuated varifocal optical element. In the past decade, metasurfaces \cite{Zheludev2012,Yu2014,Kamali2018} emerged as a versatile and compact platform for light wavefront engineering and imaging. Metasurfaces enabled the next generation of flat optics \cite{Yu2014,Lin2014,Lawrence2020}: polarizers \cite{Zhao2011,BalthasarMueller2017,Bosch2019}, efficient diffraction gratings \cite{Ding2018,Fan2018b}, holograms \cite{Ni2013,Zheng2015,Wang2016}, and lenses \cite{Khorasaninejad2016,Khorasaninejad2016a,Paniagua-Dominguez2018}. Metalenses \cite{Khorasaninejad2017} can deliver the functionalities of conventional convex or concave lenses within extremely thin form-factors by virtue of the engineered phase response of nanoscale plasmonic \cite{Fu2010} or dielectric \cite{Wang2018,Shrestha2018} resonators. However, the focal distance of a metalens is usually pre-determined by nanofabrication, and special measures must be taken to enable varifocal functionalities. Varifocal metalenses have been shown by stretching the substrate \cite{Kamali2016a}, changing the polarization of the incoming beam \cite{Zhou2020}, inhomogeneous heating \cite{Afridi2018} and incorporating voltage-controlled elements \cite{Chen2017}. Electrically actuated metalenses represent one of the most attractive varifocal devices with vast potentials for integration with existing head-up displays and lightweight cameras. Liquid crystals (LCs)---ordered assemblies of highly anisotropic molecules---are ubiquitous in modern technology as highly efficient electrically actuated switching agents. Applying AC voltage across a liquid crystal cell (LCC) can change the alignment of LC molecules, enabling high-contrast refractive index tuning. Sensitivity of metasurface resonances to the refractive index of the surrounding medium has been used to show LC-enabled electrically and thermally switched resonances \cite{Muller2002,Kossyrev2005,Gorkunov2008,Xiao2009,Kang2010,Decker2013,Buchnev2013,Atorf2014,Sautter2015,Buchnev2015,Komar2017,Parry2017,Xie2017}, absorption \cite{Shrekenhamer2013,Wang2017}, beam deflection \cite{Komar2018,Chung2020,Gorkunov2020} and aberrations \cite{Shen2020}, as well as programmable spatial light modulation \cite{Li2019,Wu2020}. While electrically actuated ultrathin optics would advance the development of compact imaging systems, LC-based varifocal metalenses have remained elusive so far.

In this Letter, we demonstrate an ultrathin LC-based metalens whose focal length can be adjusted by voltage. We provide a generalized design approach for the constituent elements of the metalens, silicon-based meta-atoms, which have optimized phase responses for several intermediate orientations $\theta$ of the LC molecules. By generating a vast meta-atom library, imposing a set of appropriate constraints, and choosing optimized meta-atom candidates, we construct a metalens that shows continuous focal distance shifting from $f_{\rm off}=15$~mm to $f_{\rm on}=12$~mm as a function of $\theta$ at a wavelength of $\lambda=690$~nm. We extend the LC-driven metalens concept to conceive two wide-aperture bifocal lens with focal shifts as large as from $f_{\rm off}$ to $f_{\rm on}=f_{\rm off}/2$ and $f_{\rm on}=-f_{\rm off}$, respectively. We validate this concept by designing and fabricating a bifocal metalens that emulates a transition between a concave and convex lens. Energy redistribution between the $f_{\rm off}$  and $f_{\rm on}=-f_{\rm off}$ focal spots manifests itself as intensity switching by up to 55\%, driven by low LCC voltages of $<10$~V$_{\rm pp}$. Ultrathin LC-driven metalens show flexible voltage-adjustable focal length control, appealing for use in compact and lightweight imaging devices.

\section{Results and discussion}

\begin{figure*}
\includegraphics[width=1\textwidth]{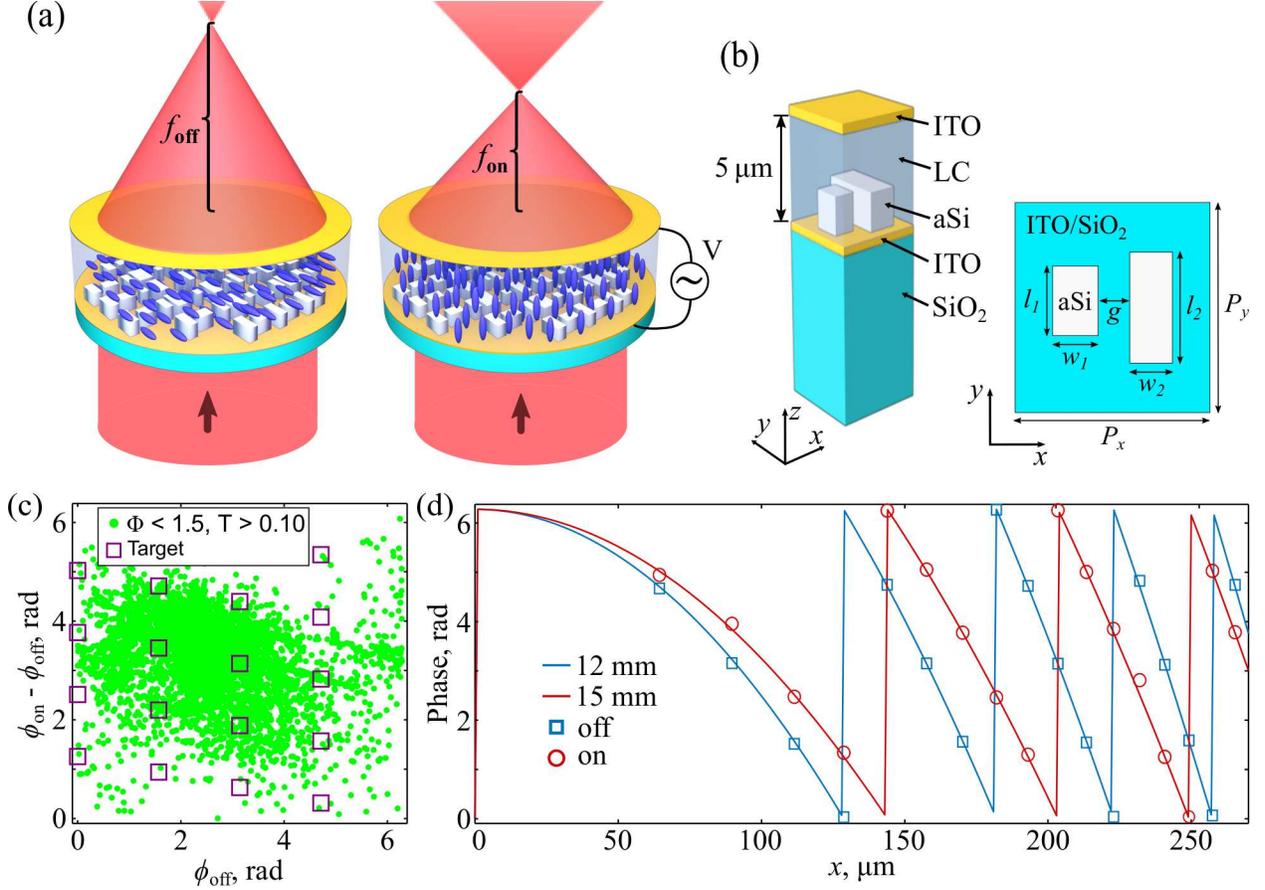}
\caption{\textbf{An ultrathin varifocal metalens encapsulated in electrically-biased liquid crystal cell (LCC).} (a) The device and its operating principles: the focal length continuously varies from $f_{\rm off}$ (no bias) to $f_{\rm on}(V_0)$ (AC bias: $V=V_0 \cos 2\pi \nu t$). (b) Unit cell schematic: an array of amorphous silicon (aSi)-based meta-atoms encapsulated in an LCC is sandwiched between two biasing indium-tin-oxide (ITO) electrodes. (c) Green dots: meta-atom candidates selected from the five-dimensional parameter space $(l_{1,2},w_{1,2},g)$ satisfying the phase-uniformity  and transmittance criteria: $\Phi<1.5$ (see Eq.(\ref{eq:phi})) and $T > 0.10$. Red squares: target values of $\phi_{\rm on}$ and $\phi_{\rm off}$ (four in each Fresnel zone). (d) The resulting phase distribution of the $4$-zone varifocal metalens in the `off' (blue squares) and `on' (red circles) states. Solid lines: the ideal phase functions of converging lenses with $f_{\rm off}=15$~mm and $f_{\rm on}=12$~mm. Other parameters: metasurface height $h=300$~nm, lattice periodicities: $P_x = P_y=405$~nm}\label{fig:schematic}
\end{figure*}

\textbf{Varifocal metalens concept}
The concept of the electrically-actuated metalens is depicted in Fig.~\ref{fig:schematic}. It belongs to the family of Fresnel lenses: the concept that has been known and widely used since early $19$th century. Because of their inherently thin nature, Fresnel lenses have been installed (and are still found) atop of lighthouses, where low-weight lenses are particularly attractive. A conventional lens focuses light at a focal distance of $f$ using a continuously-varying phase delay profile $\phi(r,f)$ given by~\cite{Goodman2005}:
\begin{equation}
\phi(r,f) =\frac{2\pi}{\lambda} (f-\sqrt{r^2 + f^2}) ,
\label{eq:f}
\end{equation}
where $r$ is the radial distance from the lens center. Because a continuously-varying phase delay requires that the lens thickness differential between its center and periphery exceeds the wavelength of light $\lambda$, Fresnel lenses overcome this limitation by dividing the lens into $m>1$ Fresnel zones, each of which produces phase delay variations in the $0 < \phi_m < 2\pi$ ranges. The phase is ``reset'' at the beginning of each Fresnel zone. 

A Fresnel metalens advances this idea further by discretizing each zone into $n$ sub-zones, with each sub-zone producing constant phase delays $0 < \phi_{m,n} < 2\pi/n$. Each sub-zone is comprised of an array of amorphous silicon (aSi) nanopillars, referred to as meta-atoms in the rest of this paper, encapsulated in a nematic LC. The design goal is to engineer the meta-atoms so as to impart a spatially-varying phase delay profile $\phi_{m,n}$ onto the transmitted light, and to control $\phi_{m,n}\equiv \phi_{m,n}(\theta)$ using the rotation angle $\theta$ of the surrounding LC molecules. %(see Fig.~\ref{fig:varifocal}). 
By varying the amplitude $V_0$ of an AC voltage $V=V_0 \cos 2\pi \nu t$ applied to the LCC, the LC orientation angle $\theta \equiv \theta(V_0)$ can be varied to ensure that the discretized phase delay  $\phi_{m,n}(\theta)$ approximates $\phi(r,f(V_0))$ given by Eq.~(\ref{eq:phi}), where $f(V_0)$ is the voltage-dependent focal length. By varying bias voltage between its ``off'' ($V_0=0$) and ``on'' ($V_0=V_{\rm max}$) values, the focal distance $f_{\rm on} < f(V_0)< f_{\rm off}$ varies correspondingly. 

A schematic of the aSi meta-atom, each comprised of two rectangular aSi pillars on an ITO-coated fused silica substrate, is shown in Fig. 1(b). Such meta-atom structures support localized electric and magnetic Mie-type resonant modes which may be spectrally tuned by modifying the permittivity of the surrounding media. Dielectric metasurfaces with moderately-high quality factors of the order $10 < Q < 100$ have been used for various applications, including polarization conversion and nonlinear optics~\cite{wu_NatComm14,Bosch2019, Neuner2013, Shcherbakov2019}. %   {\bf and maybe Max's N.Comm., maybe a few more}. 
The double-pillar meta-atom geometry shown in Fig. 1(b) enables flexible control over the transmittance and the phase by varying the lengths $l_{1,2}$ and the widths $w_{1,2}$ of the pillars, as well as the gap $g$ between them and the array periods $P_x$ and $P_y$.

The metasurface is covered by a 5-$\mu$m-thick LC and ITO-coated fused silica top plate. The ITO layer renders the top and bottom glass plates conductive for use as electrodes that apply an AC bias across the LCC. The LC surrounding the metasurface can then be modeled as an anisotropic dielectric medium with $n_{\rm o}$ = 1.51, $n_{\rm e}$ = 1.72, and $V_0$-dependent optical axis orientation angle $\theta$ (see Supporting Information Section 1 and Fig.2(a) for the definition of $\theta$). In absence of an external voltage ($V_0 = 0$), the brushed top electrode  induces a preferred in-plane anchoring direction of the LC along $y$ across the whole LCC \cite{Komar2017}; equivalently, here, $\theta$ = 0$^\circ$. Applying a bias voltage with amplitude $V_{0}>0$ can increase $\theta$ up to a maximum value of $90^\circ$, therefore modifying the phase response of the meta-atom.

In contrast to some metasurface tuning strategies which tune the entire metasurface homogeneously to modify global transmission or absorption properties \cite{Pryce2010, Yao2014, Wang2015, Ou2013,dabidian_ACSPhot14,dabidian_NanoLett16}, adjusting the focusing profile of a metalens requires the phase modulation function to be tailored locally for each meta-atom. In order to fulfill this requirement, it is critical for the aSi meta-atom design space to permit both initial imparted phases $\phi_{\rm off}$ and phase gradient $\Delta \phi \equiv \phi_{\rm on}-\phi_{\rm off}$ variations of the meta-atoms spanning the $0 < \Delta \phi < 2\pi$ range, while also maintaining acceptable transmission amplitudes. We find this prerequisite is achieved for meta-atom thicknesses in the range $0.4 \lambda < h < 1.5\lambda$, where $\lambda$ is the targeted operation wavelength. Because fabricating meta-atom arrays of variable thickness using standard top-down methods is challenging, we select the same thickness $h=0.4\lambda$ for all meta-atoms.

\textbf{Design principle}
To approximate $\phi(r,f(\theta))$, we design a Fresnel-type metalens discretized into a series of concentric rings: Fresnel zones. Each ring (zone) corresponds to a $2\pi$ phase increment of the lens, and is discretized by at least $n$ = 3 meta-atom phase steps (sub-zones). To design an $m$-zone metalens with $f=f_{\rm off}$ in the `off' state, $n$ aSi meta-atoms per zone with varying geometric parameters are selected to impart phase shifts from $0 < \phi_{m,n}(\theta=0^{\circ}) < 2\pi$ in increments of $2\pi/n$, and arranged to approximate the spatial phase profile according to $\phi_{m,n}(\theta=0^{\circ}) \approx \phi(r, f_{\rm off})$.
To enable electrical control over the focal length, the same procedure is followed in choosing the $N=n*m$ total meta-atoms to simultaneously approximate the desired phase profile according to $\phi_{m,n}(\theta=90^{\circ}) \approx \phi(r, f_{\rm on})$ for a LC molecule orientation of $\theta = $~90$^\circ$, corresponding to $V_0 = V_{\rm max}$.%reworded

 This selection enables a \emph{discrete} bifocal lens switch: tuning the LC molecule orientation from 0$^\circ$  to 90$^\circ$  switches the focal distance of the metalens from $f_{\rm off}$ to $f_{\rm on}$. While discrete switching operation may be attractive in a variety of applications\cite{Love2009,Li2006}, a truly varifocal metalens design is contingent on the meta-atoms satisfying two additional criteria at the intermediate LC molecule rotation angles $0^{\circ} < \theta < 90^{\circ} $. 

Using numerical simulations, we found that imposing these two criteria over five discrete values of   $\theta=0^\circ, 22^\circ, 45^\circ, 67^\circ$ and $90^\circ$, provides a good balance between the density of the candidate library and the quality of the varifocal operation. First, the phase for each meta-atom is sought as a monotonic (decreasing, in our case of $f_{\rm off} >f_{\rm on}$) function of $\theta$.  Therefore, the phases of the meta-atoms $\phi(\theta)$ at the five values of $\theta$ have been set to satisfy the following constraints: $\phi(90^\circ)<\phi(67^\circ)<\phi(45^\circ)<\phi(22^\circ)<\phi(0^\circ)$. Second, we found that the uniformity of the spacing between the phases $\phi(\theta)$ on the segment of $\theta\in [0^\circ,90^\circ]$ is of paramount importance in obtaining a truly varifocal lens (see Supporting Information Section 2). In other words, the preferable design requires that $\phi(\theta)$ changes near-linearly with the LC orientation angle $\theta$. For quantitative characterization, we introduce the uniformity parameter $\Phi$ defined as
\begin{equation}
\Phi = \sum_{i=1}^3 \frac{|\Delta\Phi_{i+1}| - |\Delta\Phi_{i}|}{3} ,
\label{eq:phi}
\end{equation}
where $\Delta\Phi_{1} = \phi(90^\circ) - \phi(67^\circ),
\Delta\Phi_{2} = \phi(67^\circ) - \phi(45^\circ),
\Delta\Phi_{3} = \phi(45^\circ) - \phi(22^\circ),$ and $\Delta\Phi_{4} = \phi(22^\circ) - \phi(0^\circ)$. %, and $\overline{\Delta\Phi} = \frac{1}{4}\sum_{i} \Delta \Phi_i$. 
We found that applying the constraint of $\Phi < 1$~rad is important to achieve the truly varifocal function of the metalens. These two conditions ensure the selection of meta-atoms that provide a continuous and uniformly tuned phase profile of the metalens with $V_0$. 
%{\bf Melissa: I am a bit sceptical about this dependence on $\theta$. After all, $\theta$ is just a parameter, what difference does it make how $\phi(\theta)$ depends on it as long as it is monotonic? I think what matters is the dependence of $\Delta \Phi_i$ on $\phi_i$.}

To illustrate the LC-encapsulated meta-atom optimization and selection process for a varifocal metalens, Fig. 1(c) plots a map of simulated candidates which satisfy all the imposed restrictions as a function of the two key phase quantities of a tunable lens, $\phi_{\rm off}$ and $\Delta \Phi \equiv \phi_{\rm on} - \phi_{\rm off}$. Individual green points correspond to aSi unit cells with unique geometric parameters from a five-dimensional parameter space $(l_{1,2},w_{1,2},g)$. The phase map covers almost the full 0--2$\pi$ range along both axes, thereby permitting a wide range of $f_{\rm on}$ and $f_{\rm off}$ combinations. For example, squares on the map represent the targeted phase points for a lens with $f_{\rm off} = 15$~mm,  $f_{\rm on}=12$~mm, $n=4$ sub-zones, and $m=4$ Fresnel zones. The overlap between squares and green points in Fig.~\ref{fig:schematic}(c) signifies the existence of at least one specific meta-atom candidate satisfying the metalens design. 
%{\bf I also see a bunch of blue dots in (c). What do they mean? Also, I understand that the horizontal spacing between the squares is $\pi/n$. What about the vertical one? Is it the same}. 
The selected candidates are plotted in Fig. 1(d) as a function of $x$, arranged according to Eq.(\ref{eq:f})  and showing close agreement with the `off' and `on' targeted phase profiles.

\begin{figure*}
\includegraphics[width=1.0\textwidth]{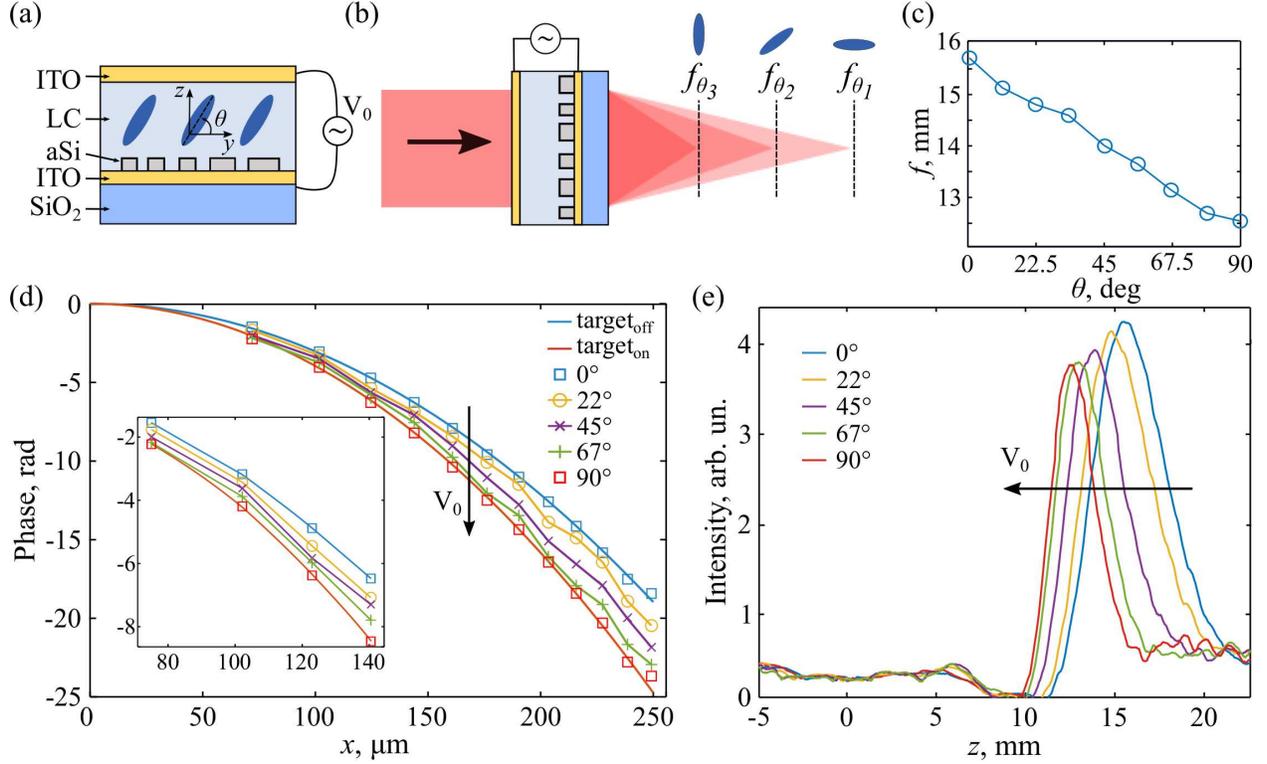}
\caption{ \textbf{Liquid-crystal-driven metalens with a continuously tunable focal spot.} (a) Schematic of the device. A silicon-based metalens is encapsulated in an LCC with a thickness of 5~$\mu$m between two transparent conducting oxide electrodes. An AC voltage is applied to the electrodes, driving the orientation of the LC molecules at angle $\theta$ with respect to the rubbing direction ($y$). (b) Applying voltage to the electrodes controls $\theta$, continuously shifting the position of the focal spot. $\theta_{1,2,3}$ denote three representative LC angles with $\theta_1<\theta_2<\theta_3$, corresponding to three distinct focal spots. c) Focal length of the lens as a function of $\theta$. Solid line serves to guide the eye. (d)  The phases of the scattered secondary waves by meta-atoms at positions $x$ controlled by the LC in a cylindrical lens implementation. %{\bf Why $x$, not $r$? Is this a cylindrical lens?}. 
(e) On-axis intensity along the optical axis $z$ for five values of $\theta$.}\label{fig:varifocal}
\end{figure*}

\textbf{Computational design of a varifocal metalens.}
An LC-driven metalens with continuously tunable focal length is depicted in Fig2.(a,b). % {\bf There are some colors in (b); what do they signify? Also, $\theta_{1,2,3}$ are not defined.}
Owing to the voltage-induced realignment of the LC molecules to angles $\theta$ with respect to the `off' ($y$: see Fig.2(a)) direction, applying an AC voltage bias across the LCC continuously controls the focal spot position. 

A three-zone ($m=3$) metalens template with $\lambda_0$ = 690~nm, $\Phi = 1$~rad, width of 500~$\mu$m, and focal distance targets $f_{\rm off}=~$12~mm and $f_{\rm on}=~$15~mm is designed to numerically demonstrate the focusing performance and varifocal tuning capability of a LC-driven metalens. We realize full-aperture simulations of the varifocal metalens by implementing the design as a cylindrical lens. %
The resulting simulated focal length of the metalens is plotted in Fig. 2(c) as a function of $0^\circ < \theta < 90^\circ$. A smooth and near-constant rate of change of $f$ with the LC angle is observed: $f^{-1}(\Delta f/\Delta \theta)\approx 0.22\%\cdot$deg$^{-1}$. The phase profile imparted onto transmitted light by the cylindrical metalens as a function of $x$ is presented in Fig. 2(d) for five values of $\theta$. These results demonstrate a uniform and continuous phase modulation of twelve total sub-zones by modulating the LC orientation angle. The $x-$dependent phase imprinted by the metalens onto normally incident light starts near the targeted $\phi_{\rm off}(r,f_{\rm off} = 15~\mbox{mm})$ phase profile at $\theta=0^\circ$ (`off' orientation), and ends near the targeted $\phi_{\rm on}(r,f_{\rm on} = 12~\mbox{mm})$ phase profile at $\theta=90^\circ$ (`on' orientation). Remarkably, each of the five values of $\theta$ produces high-contrast focal spots along the optical axis of the lens ($z$) as depicted in Fig.~2(e). The focusing efficiency of the metalens, determined by the ratio of total focal plane intensity in an area with three times the full width at half-maximum (FWHM) of focal spot to the total intensity of the incident beam \cite{Arbabi2015}, is found to be 12.1$\%$ for the `off' LC orientation and and 13.6$\%$ for `on' LC orientation (see Supporting Information, Section 1).

While $f_{\rm off}=~$15~mm and $f_{\rm on}=~$12~mm were the targets for our numerical demonstration, we note that the adaptability of our optimization approach enables the design of varifocal metalenses with nearly-arbitrary focal length tuning ratios $T$, defined as $T = \frac{f_{\rm off}-f_{\rm on}}{f_{\rm off}}$ \cite{Cheng2006,Tseng2018}. %{\bf Would be good to give a reference for this definition, if available}. 
We found that imposing the varifocal phase restrictions over five discrete $\theta$ values is generally sufficient to facilitate the design of varifocal metalenses with $T<$~0.3, whereas varifocal operations with $T>$~0.3 are accessible through imposing the phase uniformity condition over a finer-range of discrete $\theta$-values, i.e. $\theta\in[0^\circ:\frac{90^\circ}{k}:90^\circ]$ with $k>~$5.

\textbf{Simplified computational design of large aperture bifocal metalenses}
Several modern imaging technologies require tunable focus lenses with large apertures \cite{Hong2017}. Scaling the metalens size can pose optimization challenges, since typically each 2$\pi$ zone added to the lens requires $n$ (one per sub-zone) unique meta-atom designs. As an example, a 1-mm-wide lens with $f_{\rm off} =~$15~mm, $f_{\rm on}=~$12~mm, $\lambda=~$690~nm, and $P_x = P_y =~$470~nm, requires approximately 100 unique meta-atoms, possibly pushing the limits of the library or fabrication routines.

Here, we propose several types of LC-metalenses with judiciously selected focusing conditions for two discrete focal spots to bypass the scaling limitations. Figures 3(a,b) present two proposed types of bifocal LC-metalenses that can be scaled to large aperture sizes. For Type I, the focus changes from $f_{\rm off}$ to $f_{\rm on} = f_{\rm off}/2$, and for Type II, the focus changes from $f_{\rm off}$ to $f_{\rm on} = -f_{\rm off}$ upon changing the orientation of the LC molecules from $\theta=0^\circ$ to $\theta=90^\circ$. The reduced functionality of the bifocal metalens enables a simplified metalens design with just three ($n=3$) types of engineered meta-atoms. Nevertheless, this small number of designer meta-atoms is sufficient for creating high-performance LC-tunable bifocal metalenses with nearly-arbitrary size.

\begin{figure*}
\includegraphics[width=1\textwidth]{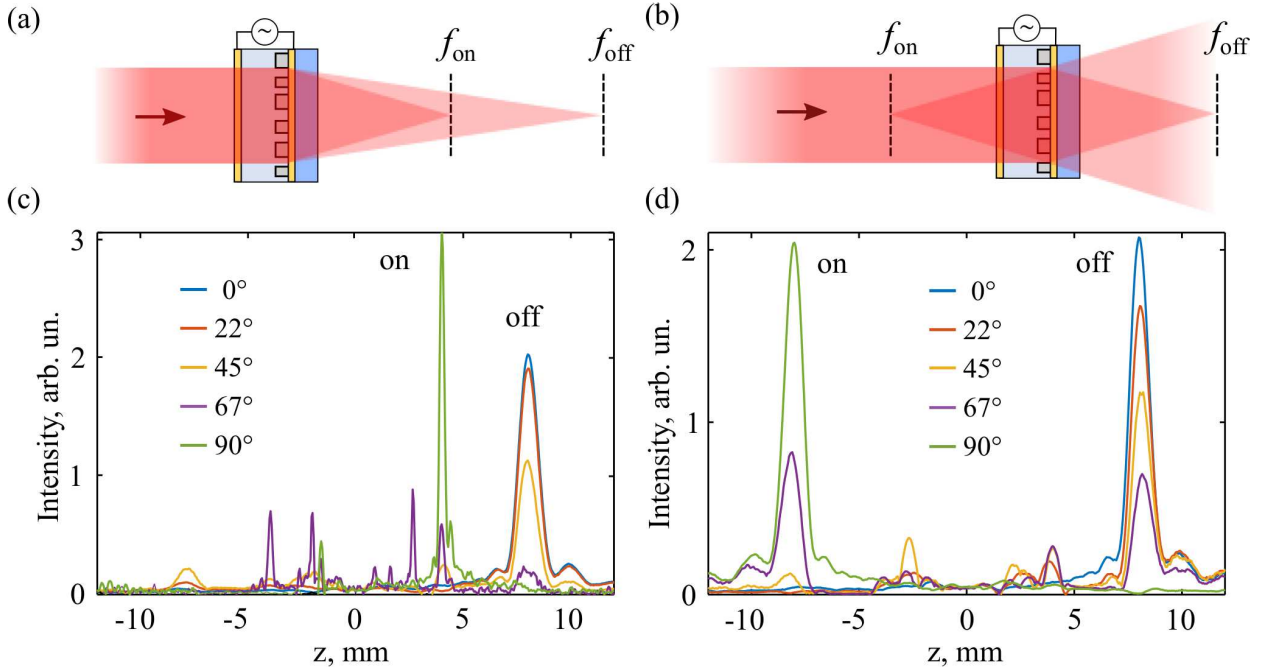}
\caption{\textbf{Simulation of a bifocal LC-controlled cylindrical metalens.} (a) A sketch of Type I bifocal metalens: upon changing the orientation of the LC molecules from $\theta=0^\circ$ to $\theta=90^\circ$, the focal length of the metalens changes from $f_{\rm off}$ to $f_{\rm on} = f_{\rm off}/2$.  (b) A sketch of Type II bifocal metalens: upon changing the orientation of the LC molecules from $\theta=0^\circ$ to $\theta=90^\circ$, the focal length of the metalens changes from $f_{\rm off}$ to $f_{\rm on} = -f_{\rm off}$.  (c) Intensity along $z$ showing focusing at distances from $f_{\rm off}=8$~mm ($\theta=0^\circ$) to $f_{\rm on}=4$~mm ($\theta=90^\circ$) for a Type I bifocal metalens. (d) Intensity along $z$ showing focusing at distances from $f_{\rm off}=8$~mm ($\theta=0^\circ$) to $f_{\rm on}=-8$~mm ($\theta=90^\circ$) for a Type II bifocal metalens.}\label{fig:bifocal_cylindrical}
\end{figure*}

Numerical demonstrations of wide-aperture LC-driven bifocal metalenses are accomplished for lens widths of 1~mm, $\lambda_0=~$808~nm, $n=5$, and $f_{\rm off}=8$~mm by applying the standard design routine and disregarding the additional varifocal constraints. The simulated intensity of light along the optical axis of the Type I and Type II cylindrical metalenses are plotted in Figs. 3(c,d), demonstrating high-contrast switching from $f_{\rm on}=8$~mm to $f_{\rm off}=4$~mm and $f_{\rm on}=8$~mm to $f_{\rm off}=-8$~mm, respectively, when the LC alignment is changed from $\theta=0^\circ$ to $\theta=90^\circ$. 
The corresponding calculated focusing efficiencies of the Type I and Type II metalenses range between 22-27$\%$ $\theta=0^\circ$ and $\theta=90^\circ$ (See Supporting Information Section 1). The focusing profiles of the lenses at intermediate LC molecule angles reveal the optical intensity is redistributed gradually between the two focal spots as a function of $\theta$. The smaller peaks appearing at integer fractions of $f_{\rm off,on}$ are expected due to the discretization of the phase profile by the metalens, and their relative intensities can be diminished by increasing the number of different meta-atoms ($n$) used.

\textbf{Experimental demonstration of a LC-controlled cylindrical bifocal metalens}
The proof-of-concept experimental demonstration focuses on a Type II bifocal LC metalens design template consisting of single-pillar aSi meta-atoms, i.e  $w_1 = w, l_2=l, g = w_2 = l_2 =0$, and $P_x=P_y\equiv P$. To ease fabrication complexity, it was necessary for the constituent meta-atoms of the metalens to have filling factors, defined as $F = w_x w_yP^{-2}$, in the range $F=~$0.35--0.4. Heuristically, the lateral geometry and bifocal phase constraints were concurrently achieved for a Type I template with $n=~$3 sub-zones, $t=300$~nm, $\lambda=800$~nm, and $f_{\rm off} = -f_{\rm on}=~$8~mm.
 To experimentally verify the LC-driven metalens concept, a 760-$\mu$m-wide cylindrical metalens was fabricated on an ITO-coated SiO$_2$ substrate and encapsulated in a 5~$\mu$m-thick LC (see Supporting Information Sections 3 and 4). Two devices, Lens 1 and Lens 2, were fabricated. Fig. 4(a) presents the scanning electron micrograph (SEM) of Lens 2, color-coded according to the three meta-atom geometries. The imaging setup depicted in Fig. 4(b) was used to characterize the focusing and tuning performance of the samples. A $y$-polarized $\lambda=808$~nm laser diode beam was sent onto the metasurface at normal incidence, and a CMOS camera was used to image the focal spot of the metalens in transmission through a refocusing lens with a focal distance of 25.4~mm. A hybrid piezoelectric-mechanical stage enabled the translation of the metalens along the optical axis for the imaging of different $xy$-planes over a wide range of $z$-positions and AC voltages (see Supporting Information Section 6). The results of this measurement on Lens 1 is presented in Fig. 4(c), showing intensity maxima that occur at focal distances $z_{\rm } = -$ 8.6$~$mm and $z = $+8.7$~$mm for $V_0=0$~$V_{\rm pp}$ and $V_0=3.2$~$V_{\rm pp}$ respectively, as expected for the Type II switching behavior. The reversed Type-II switching ($f_{\rm off} = -f, f_{\rm on} = +f$) and its diminished contrast can be reproduced by numerical simulations which account for discrepancies between the designed and experimental metalens geometrical parameters and laser wavelength (see Supporting Information Section 7). Although it is not contained in our model, an additional factor contributing to the low switching contrast of Lens 1 could be the misalignment of the initial LC anchoring direction, a known issue of LC-metadevice fabrication \cite{Sun2019}.
 
 The focal spot intensities at $z = -f$ and $z = +f$ are plotted as a function of voltage for Lens 1 in Fig.4(d), showing that the `on' voltage, $V_0=3.2$~$V_{\rm pp}$  corresponds to the point of maximum off/on switching contrast for both focal spots. Figs. 4(e,f) present the measured voltage-dependent intensity line-cuts of the focal spots and 2D focal spot profiles of our best-fabricated and (consequently) best-performance metalens, labeled Lens 2. Lens 2 focuses light parallel to the $y$-axis per the cylindrical design. Intensity variations along $y$ of the focal lines are ascribed to the asymmetric distribution of the incident laser. By increasing the voltage from  $V_0$ =0 to  $V_0 =3.2V_{\rm pp}$, the focal spot intensity at $z = +f$ (Fig. 4(e)) increases by 55$\%$; for the same voltage increment, the focal spot intensity at $z = -f$ (Fig. 4(f)) decreases by  31$\%$. The FWHM of the $f_{\rm on}$ and $f_{\rm off}$ focal spots are 17.6~$\mu$m and 10.4~$\mu$m respectively, comparable to that enabled by the diffraction limit of 9.3~$\mu$m.% {\bf Some explanation is needed for why the two lenses perform differently. Also, why is there so much variation in $y$ shown in the two insets if this is a cylindrical lens? Need to at least comment.}.
\begin{figure*}
\includegraphics[width=0.8\textwidth]{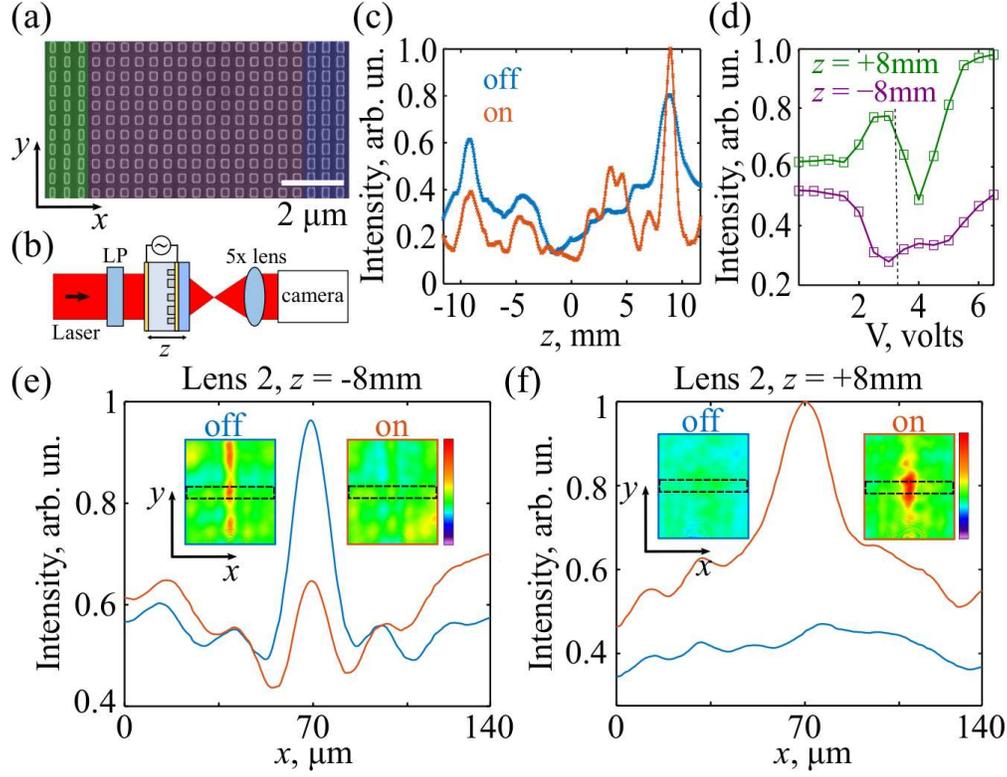}
\caption{\textbf{Experimental realization of a LC-controlled bifocal cylindrical metalens.} (a) A SEM image of a typical cylindrical metalens sample, showing a three-subzone structure. (b) Schematic of the measurement setup. LP is a linear polarizer. (c) A $z$-coordinate scan of the on-axis intensity for `off' (blue data points) and `on' (orange data points) states of Lens 1. Focal points at $x=\pm8$~mm are seen to exchange intensity as the voltage is changed between $V_0 =0$ and $V_0=3.2V_{\rm pp}$. (d) The focal spot intensities at $z=f_{\rm on}$ and $z=f_{\rm off}$ plotted as a function of $V_{0}$ for Lens 1. The dotted line indicates the `on' voltage, $V_0=3.2V_{\rm pp}$. (e) The $x$-sections of the $z=+8$~mm focal line image at $V_0=0$ (off, no distinct focal line visible) and $V_0=3.2V_{\rm pp}$ (on, focal line present). The insets show the respective camera images with the dashed rectangles denoting the $y$-averaging areas.  (f) Same as (e) but for $z=-8$~mm. Lens 2, shown in (e) and (f), shows considerably larger LC-driven modulation than than of Lens 1.
}
\end{figure*}

\textbf{Outlook and Conclusions}
Although the operation wavelengths used in this work were 690~nm and 808~nm, the spectral range can be extended through appropriate meta-atom scaling and selection of constituent meta-atom materials with low Ohmic losses. For instance, GaP \cite{Cambiasso2017} and TiO$_2$ \cite{Khorasaninejad2017, Devlin2016} or crystalline Si \cite{Miller2017,Yu2015} can enable high performance LC-driven varifocal metalenses in the visible or infrared spectral ranges, respectively. %%%*** Indeed, our preliminary simulations using $\lambda$ =1650nm and aSi as the meta-atom material suggest an ease of varifocal metalens implementations with transmission efficiencies $\epsilon>0.4$ or higher. % (see Supporting Information).
Moreover, the experimental focusing efficiency is expected to increase with improved nanofabrication and LC alignment. Finally, by integrating our platform with recent technical advances in metasurfaces \cite{Shrestha2018,Wang2018,Hu2019} we envision prospects for RGB operation and optical aberration correction, to eventually find use in compact imagers, such as head-up displays and augmented reality glasses.

To conclude, we have demonstrated the electrostatic actuation of a LC-encapsulated semiconductor metasurface for continuous and reversible varifocal lensing in the visible and near-IR spectral ranges. We present a metasurface design consisting of resonant aSi nanopillars encapsulated in a nematic LCC. A voltage bias applied across the cell tunes the resonance frequency of the aSi nanopillars, thereby enabling real-time control over the spatial phase profile imparted on the incident light and the resulting focal length of the metalens. Our numerical simulations have verified several metalens designs, including a metalens with the focal length tuning continuously from $f=12$~mm to  $f=15$~mm, as well as metalenses with the focal length switchable between two discrete focal lengths. We experimentally demonstrate a LC-driven metalens which facilitates focal length switching between $-8$~mm and $+8$~mm when a low voltage bias of 3.2~$V_{\rm pp}$ is applied across the cell. Our results establish LC-driven resonant dielectric metasurfaces as a versatile platform for electrically-actuated varifocal metalenses, suitable for various applications where compact and tunable focusing devices are sought for.

\section{Acknowledgments}
This work was supported by Global Research Outreach program of Samsung Advanced Institute of Technology.   This work was performed in part at the Cornell  NanoScale  Science  and  Technology  Facility  (CNF),  a member of the   National   Nanotechnology   Coordinated Infrastructure (NNCI), which is supported by the National Science Foundation (Grant No. NNCI-2025233). MB is supported by the National Science Foundation Graduate Research Fellowship (Grant No. DGE-1650441). Any opinion, findings, and conclusions or recommendations expressed in this material are those of the authors and do not necessarily reflect the views of the National Science Foundation

\section{Methods}
See Supporting information for the list and description of methods.

\providecommand{\latin}[1]{#1}
\makeatletter
\providecommand{\doi}
{\begingroup\let\do\@makeother\dospecials
	\catcode`\{=1 \catcode`\}=2 \doi@aux}
\providecommand{\doi@aux}[1]{\endgroup\texttt{#1}}
\makeatother
\providecommand*\mcitethebibliography{\thebibliography}
\csname @ifundefined\endcsname{endmcitethebibliography}
{\let\endmcitethebibliography\endthebibliography}{}

\newpage

\textbf{\Large Supporting Information}

\subsection{1. Simulations}
%lattice constant
The optical response of LC-encapsulated aSi-metasurfaces was simulated using the FEM software package COMSOL Multiphysics. For single-cell simulations, aSi metasurfaces were periodically-arranged single- or double-pillar meta-atoms 300 nm in height, with lattice constants $P_x,P_y$ given in Tables S1-S4, and variable pillar lengths and widths. The operation wavelength $\lambda$ is chosen to be 690~nm for varifocal metalens designs and 808~nm for bifocal metalens designs. The nematic LC surrounding the metasurface was modeled as an anisotropic dielectric medium with $n_o = 1.51$ and $n_e = 1.72$. The LC permittivity $\epsilon_{\rm LC}(\theta)$  for the `off' voltage was modeled by the diagonal tensor $\epsilon_{\rm LC}(\theta=0^\circ) = \mbox{diag}[n_o^2, n_e^2, n_o^2]$. To simulate the effect of an applied electric field and LC molecule reorientation, the LC permittivity tensor was modified to $\epsilon_{\rm LC} (\theta) = R_z^{-1}(\theta)\epsilon_{\rm LC}(\theta=0^\circ)R_z(\theta)$ where $R_z(\theta)$ is the rotation matrix about the $z$-axis. The substrate is silica coated by a 150-nm-thick ITO layer, with respective refractive indices of 1.45 and $1.6+0.01i$, respectively. The refractive index of the aSi pillars was taken to be the ellipsometry-characterized value of our PECVD-films, $3.8+0.01i$. Periodic boundary conditions were enforced along the $x$ and $y$ domain walls, and perfectly matched layers were applied to the domain boundaries parallel to the $xy$-plane. The incident field was a normally-incident plane wave polarized along the  anisotropy ($y-$)axis of the LC in the `off' state. The transmitted electric field was simulated for various meta-atom geometries and for $\theta$ ranging from from 0$^\circ$ to 90$^\circ$ in steps of 22.5$^\circ$ to generate an extensive library of meta-atom candidates.
The cylindrical bifocal and varifocal metalenses were simulated with Floquet periodic boundary conditions along the $xz$ and $yz$ domain walls and perfectly matched layer boundary conditions at $xy$-plane domain boundaries. A perfect magnetic conductor boundary condition is applied at the ($y =0$) symmetry plane of the lens to reduce computational volume by a factor of two.  The lens is illuminated by normally-incident plane wave polarized along $y$ from the substrate side. 
The transmitted electric field distribution $E(x,y,z=0)$ is propagated to planes along the $z$-axis $E(x',y',z')$ by using a Fourier propagation integral:
$$E(x',y',z') = \int_{-\infty}^{\infty} \tilde{E}(k_x,k_y)|_{z=0}e^{-i(k_xx'+k_yy')}e^{-k_zz'} dk_x dk_y,$$ where $\tilde{E}(k_x,k_y)|_{z=0}$ represents the Fourier transform of $E(x,y,z=0)$, $k_z \equiv \sqrt{k_0^2-k_x^2-k_y^2}$, and $(x',y')$ are the spatial coordinates in the plane a distance $z=z'$ from the lens. 
%{\bf Need to provide the explicit integral used.} 
The focusing efficiency of the metalens is calculated as the ratio of intensity in the focal plane in a rectangular aperture with three times the full-width half maximum (FWHM) of focal line to the total intensity of light incident on the metalens. The FWHM of simulated and experimental focal spots are calculated by fitting the focal spot intensity $I$ at $z=f$ as a function of transverse-coordinate $x$ to a normal Gaussian distribution: $ I(x )=A+ \frac{B}{\sqrt{2\,\pi \sigma^{2}}}\,\text{exp}\left[-\frac{\left(x-x_o\right)^2}{2\,\sigma^2}\right]$, where $x_0$, $A$, $B$, and $\sigma$ and are best fit parameters. Then FWHM = $2\sqrt{2 ln2}\sigma$.\\

\section{2. Varifocal lens phase uniformity justification}
Assuming its numerical aperture small, the phase profile of a lens can be approximated by $\phi(x,f) \approx\frac{-\pi x^2}{\lambda f}$. Our design objective is a varifocal lens whose focal length varies linearly with LC angle: $f(\theta) = \frac{\delta f}{\delta\theta}\theta + f_{\rm off}$, where $\delta f = f_{\rm on}-f_{\rm off}$ and $\delta\theta = \theta_{\rm on}-\theta_{\rm off}$. Thus $\phi(x,f)$ can be rewritten as 
\begin{equation}
\tag{7} 
\phi(x, f) \propto \frac{x^2}{f(\theta)}  = \frac{x^2}{\frac{\delta f}{\delta\theta}\theta + f_{off}} =\phi(\theta) \phi(x), \end{equation} 
where  $\phi(x) =x^2,\phi(\theta) = \frac{1}{1 + A\theta},$ and $A \equiv\frac{\delta f}{\delta\theta} \cdot \frac{1}{f_{\rm off}}$,  Notably, in the limit $|T|< 0.25$, %=\frac{1}{1 + A\theta}\cdot\frac{x^2}{f_{\rm off}}
$\phi(\theta) \sim 1 - A\theta $, i.e. %Since $\frac{\theta}{\delta\theta} \leq 1$, the limit is equivalent to $T< 0.25$. %|\frac{f_{\rm on}-f_{\rm off}}{f_{\rm off}}| 
varifocal phase profiles $\phi(x, \theta)$ vary linearly with $\theta$ for designs where the total tuning ratio is below 25$\%$. Alternatively stated, to design a metalens with a continuously varying focal length, meta-atoms should exhibit approximately evenly-spaced phases $\phi_i$ with uniform LC molecule orientation angle tuning. % validating the phase uniformity criterion. 
\section{3. Metalens fabrication}

A 300-nm-thick amorphous silicon (aSi) film was deposited onto a ITO-coated fused silica substrate (University Wafer) using  plasma-enhanced chemical vapor deposition (Oxford PECVD). The aSi was treated with SurPass 3000 adhesion promoter by immersing the wafer in SurPass solution for 30~s.  HSQ 6$\%$ was spun to form a 150-nm-thick layer over the aSi, baked for 2~min at 170$^{\circ}$C, coated with DisCharge anti-charging layer, e-beam exposed at a dose of 250~$\mu$C/cm$^2$ (JEOL 9500FS), and developed in 2.5$\%$ TMAH MIF300 solution for 120~s. The pattern was transferred to the aSi layer using an inductively coupled HBr plasma reactive ion etch (Oxford Cobra). The resulting samples were characterized with a scanning electron microscope (Zeiss Ultra).

\section{4. Liquid crystal encapsulation}

A polyimide was spun onto the bare ITO-coated upper glass substrate of the LCC to create a homogeneous planar LC alignment, which yielded a low pretilt angle of about 5° after rubbing. The polyimide layer was not applied the lower LCC substrate due to the presence of the metalens. The upper and lower substrates were separated with a uniform cell thickness of 5~$\mu$m, defined by sprayed spacer balls mixed into isopropyl alcohol.
Once the cell thickness was fixed by UV-curable optical adhesive, the device was filled with liquid crystal material using capillary action. Merck E7 nematic liquid crystals with positive dielectric anisotropy $\Delta\varepsilon>0$ were used to make liquid crystal director tends to align parallel to the electric-field. Bias wires were soldered onto the upper and lower substrates using indium wire. Two connections were made to the upper substrate for care ITO electrode and one connection to the bottom metalens arrays. Optical microscope images of the resulting samples are shown in Fig. S1. The the lens first from the top and second from the left of the pixel array in Fig. S1(a) is 'Lens 2'. `Lens 1' is the second from the top and third from the left of the pixel array shown in Fig. S1(b).  
\begin{figure}[hbt!]
	\includegraphics[width=0.5\textwidth]{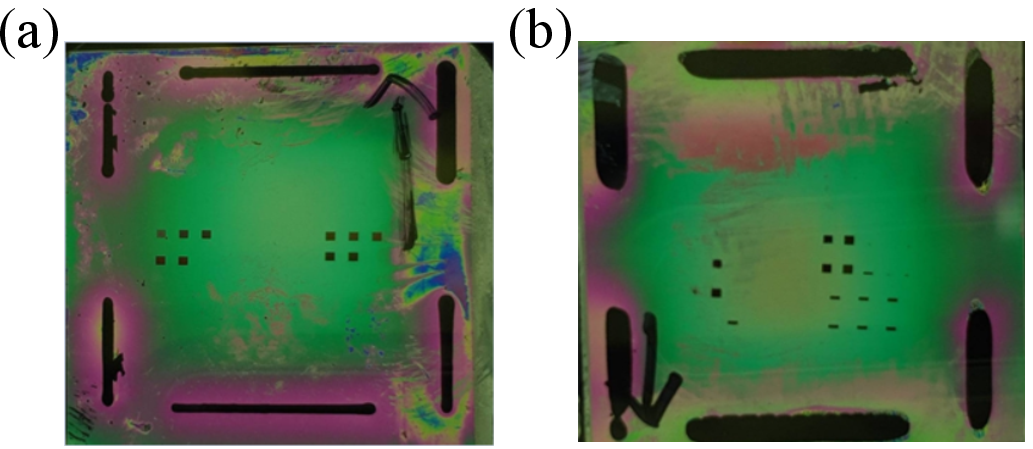}
	\caption*{Figure S1: \textbf{Optical microscope images of LC-encapsulated metalens pixel arrays}. (a) LC-encapsulated metalens array containing `Lens 2', (b) LC-encapsulated metalens array containing `Lens 1'.} 
\end{figure}

% The liquid crystal mixture of a nematic E7 provided by Merck was chosen for our LC-driven metalenses device.% and the properties of LCs exploited in the device including K11, K22 and K33 are given in Table x. The overall dimensions of the nanophotonic device are about x mm Х x mm as shown in Fig x. (Picture of an actual LC-based metalens device will be added)

\section{5. Measurements}

To characterize the focusing and tunability of the metalens, a 808~nm laser diode (Thorlabs L808P500MM) was used as a source of radiation, linearly-polarized along the LC-anchoring direction ($y$) by a wire-grid polarizer, and sent onto the metalens at near-normal incidence. % with 500$\mu$m beam-width.
Light transmitted through the metalens was magnified by a factor of 5 using a CaF$_2$ lens and the resulting intensity distribution in the $xy$-plane was imaged by a CMOS camera (Thorlabs CS165MU) coupled to a 780~nm long-pass filter. A piezoelectric stage (Newport 8303 Picomotor Actuator) enabled the translation of metalens along the optical ($z-$)axis for the imaging of different $xy$-planes. An AC bias voltage of 3.2$V_{\rm pp}$ at 1~kHz was applied to the LCC to switch between the `off' and the `on' LC molecule orientations. For both orientations, the intensity of transmitted light in the $xy$-plane was recorded as the metalens $z$-position was varied from $z = -11$~mm to 11~mm in increments of 63.5~$\mu$m, where $z=~0$~mm corresponds to the image point of the CaF$_2$ magnifying lens.

\section{6. Metalens geometric parameters}

Tables S1, S2 and S3 present the geometrical parameters of meta-atoms used for the simulations corresponding to Fig. 2, Fig. 3(a) and Fig. 3(b) of the main text, respectively. Table S4 lists the meta-atom geometries of the bifocal template used for our experimental demonstration, corresponding to Fig. 4(a) of the main text. All tabulated parameters have units in nanometers.

\begin{table}[hbt!]
	\centering
	\begin{tabular}{| c | c | c | c | c | c | c | c | c|} 
		\hline
		zone& subzone & $w_1$ & $l_1$  &  $w_2$ & $l_2$ & $g$ & $P_x$ & $P_y$ \\ 
		\hline
		1 &1 &68 &237 &0 &0 &0 &400 &385\\ \hline
		1 &2 &56 &237 &0 &0 &0 &400 &385\\ \hline %%
		1 &3 &200 &310 &106 &298 &27 &405 &405\\ \hline%%
		1 &4 &103 &289 &139 &181 & 33 &405 &405\\ \hline%%
		2 &1 &43 &250 &161 &128 &58 &405 &405\\ \hline%%
		2 &2 &174 &92 &68 &89 &54 &405 &405\\ \hline%%
		2 &3 &94 &86 &47 &71 &132 &405 &405\\ \hline%%
		2 &4 &181 &339 &0 &0 &0 &400 &385\\ \hline%%SP
		3 &1 &142 &340 &0 &0 &0 &400 &385\\ \hline%%SP
		3 &2 &46 &170 &0 &0 &0 &400 &385\\ \hline%%SP
		3 &3 &110 &48 &119 &125 &113 &405 &405\\ \hline%%
		3 &4 &55 &193 &0 &0 &0 &400 &385\\ %SP
		% 0.0679    0.2369    0.0679    0.2369    0.0679%SP
		%    0.0557    0.2372    0.0557    0.2372    0.0557%SP
		%    0.1999    0.3094    0.1062    0.2981    0.0267
		%    0.1026    0.2890    0.1389    0.1808    0.0332
		%   0.0426    0.2501    0.1608    0.1282    0.0580
		%    0.1735    0.0919    0.0678    0.0892    0.0542
		%   0.0937    0.0863    0.0472    0.0709    0.1321
		%    0.1812    0.3390    0.1812    0.3390    0.1812%SP
		%    0.1421    0.3401    0.1421    0.3401    0.1421%SP
		%    0.0456    0.1702    0.0456    0.1702    0.0456%SP
		%    0.1103    0.0479    0.1190    0.1245    0.1125
		%    0.0546    0.1929    0.0546    0.1929    0.0546%SP
		\hline
	\end{tabular}
	\caption*{Table S1: Meta-atom parameters used in the varifocal metalens simulation (Fig. 2 of the main text).}
	\label{table:1}
\end{table}

\begin{table}[hbt!]
	\centering
	\begin{tabular}{| c | c | c | c |} 
		\hline
		subzone & $w_1$ & $l_1$  &$P$ \\ 
		\hline
		1 & 124 & 224 & 475 \\ \hline
		2 & 192 & 145 & 475 \\ \hline
		3 & 229 & 322 & 475 \\ \hline	
		4 & 159 & 249 & 475	\\ \hline
		5 & 257 & 391 & 475	\\
		\hline
	\end{tabular}
	\caption*{Table S2: Meta-atom parameters used in the Type I bifocal metalens simulation (Fig. 3(a) of the main text)}
\end{table}

\begin{table}[hbt!]
	\centering
	\begin{tabular}{| c | c | c | c |} 
		\hline
		subzone & $w_1$ & $l_1$  &$P$ \\ 
		\hline
		1 & 279 & 384 & 475 \\ \hline
		2 & 283 & 359 & 475 \\ \hline
		3 & 141 & 166 & 475 \\ \hline
		4 & 125 & 420 & 475 \\ \hline
		5 & 129 & 274 & 475 \\
		\hline
	\end{tabular}
	\caption*{Table S3: Meta-atom parameters used in the Type II bifocal metalens simulation (Fig. 3(b) of the Main text)}
\end{table}

\begin{table}[hbt!]
	\centering
	\begin{tabular}{| c | c | c | c |} 
		\hline
		subzone & $w_1$ & $l_1$  &$P$ \\ 
		\hline
		1 & 272 & 382 & 470 \\ \hline  
		2 & 310 & 327 & 470 \\ \hline  
		3 & 233 & 390 & 470 \\
		\hline
	\end{tabular}
	\caption*{Table S4: Meta-atom parameters of the experimental Type II bifocal metalens template  (Fig. 4 of the main text)}
\end{table}

\section{7. Supporting calculations}

Fig. 4 of the main text presents the measured focusing and tuning of the fabricated cylindrical lens. The fabricated lenses were based on a Type-II bifocal template (as shown in Fig. 3 of the main text) for $f=8$~mm and an operation wavelength of 800~nm. To closely compare the experimental results with theory, the metalens focusing and tunability was numerically calculated using geometrical parameters matching that of the fabricated metalens and an operation wavelength matching that of the laser diode used in experiment (808~nm). Fig. S2 presents the calculated intensity of light along the optical axis as a function of $z$ and $\theta$, showing agreement with the experimental trend of increasing $z=-f$ focal spot intensity and decreasing $z=+f$ focal spot intensity for $0^\circ\leq \theta \leq 45^\circ$. The main difference between the experimental and simulated results is the relative intensities of the focal spots; possible reasons for this deviation include structural disorder of the fabricated sample and the inhomogeneous beam profile of the incident laser in experiment. 

\begin{figure}[hbt!]
	\includegraphics[width=1\textwidth]{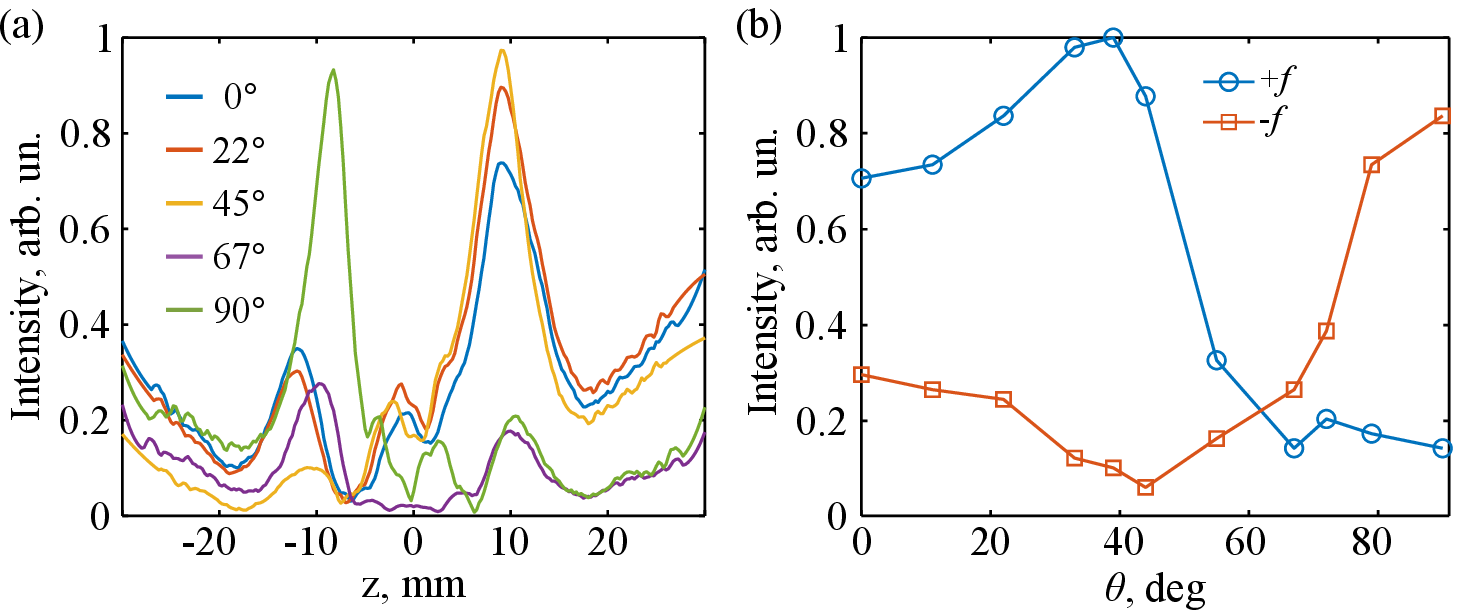}
	\caption*{Figure S2: \textbf{Simulated metalens focusing and tunability using the geometrical parameters of the fabricated metalens} (cf. Fig.4 of the main text). (a) Simulated intensity of light along the optical axis of the metalens as a function $z$. (b) Calculated intensity of the real and virtual focal spots, as a function of $\theta$ and normalized to the maximum spot intensity.} 
\end{figure}


\begin{mcitethebibliography}{68}
	\providecommand*\natexlab[1]{#1}
	\providecommand*\mciteSetBstSublistMode[1]{}
	\providecommand*\mciteSetBstMaxWidthForm[2]{}
	\providecommand*\mciteBstWouldAddEndPuncttrue
	{\def\EndOfBibitem{\unskip.}}
	\providecommand*\mciteBstWouldAddEndPunctfalse
	{\let\EndOfBibitem\relax}
	\providecommand*\mciteSetBstMidEndSepPunct[3]{}
	\providecommand*\mciteSetBstSublistLabelBeginEnd[3]{}
	\providecommand*\EndOfBibitem{}
	\mciteSetBstSublistMode{f}
	\mciteSetBstMaxWidthForm{subitem}{(\alph{mcitesubitemcount})}
	\mciteSetBstSublistLabelBeginEnd
	{\mcitemaxwidthsubitemform\space}
	{\relax}
	{\relax}
	
	\bibitem[Kildishev \latin{et~al.}(2013)Kildishev, Boltasseva, and
	Shalaev]{Zheludev2012}
	Kildishev,~A.~V.; Boltasseva,~A.; Shalaev,~V.~M. {Planar Photonics with
		Metasurfaces}. \emph{Science} \textbf{2013}, \emph{339}, 1232009\relax
	\mciteBstWouldAddEndPuncttrue
	\mciteSetBstMidEndSepPunct{\mcitedefaultmidpunct}
	{\mcitedefaultendpunct}{\mcitedefaultseppunct}\relax
	\EndOfBibitem
	\bibitem[Yu and Capasso(2014)Yu, and Capasso]{Yu2014}
	Yu,~N.; Capasso,~F. {Flat optics with designer metasurfaces}. \emph{Nature
		Materials} \textbf{2014}, \emph{13}, 139--150\relax
	\mciteBstWouldAddEndPuncttrue
	\mciteSetBstMidEndSepPunct{\mcitedefaultmidpunct}
	{\mcitedefaultendpunct}{\mcitedefaultseppunct}\relax
	\EndOfBibitem
	\bibitem[Kamali \latin{et~al.}(2018)Kamali, Arbabi, Arbabi, and
	Faraon]{Kamali2018}
	Kamali,~S.~M.; Arbabi,~E.; Arbabi,~A.; Faraon,~A. {A review of dielectric
		optical metasurfaces for wavefront control}. \emph{Nanophotonics}
	\textbf{2018}, \emph{7}, 1041--1068\relax
	\mciteBstWouldAddEndPuncttrue
	\mciteSetBstMidEndSepPunct{\mcitedefaultmidpunct}
	{\mcitedefaultendpunct}{\mcitedefaultseppunct}\relax
	\EndOfBibitem
	\bibitem[Lin \latin{et~al.}(2014)Lin, Fan, Hasman, and Brongersma]{Lin2014}
	Lin,~D.; Fan,~P.; Hasman,~E.; Brongersma,~M.~L. {Dielectric gradient
		metasurface optical elements}. \emph{Science} \textbf{2014}, \emph{345},
	298--302\relax
	\mciteBstWouldAddEndPuncttrue
	\mciteSetBstMidEndSepPunct{\mcitedefaultmidpunct}
	{\mcitedefaultendpunct}{\mcitedefaultseppunct}\relax
	\EndOfBibitem
	\bibitem[Lawrence \latin{et~al.}(2020)Lawrence, Barton, Dixon, Song, van~de
	Groep, Brongersma, and Dionne]{Lawrence2020}
	Lawrence,~M.; Barton,~D.~R.; Dixon,~J.; Song,~J.~H.; van~de Groep,~J.;
	Brongersma,~M.~L.; Dionne,~J.~A. {High quality factor phase gradient
		metasurfaces}. \emph{Nature Nanotechnology} \textbf{2020}, \emph{15},
	956--961\relax
	\mciteBstWouldAddEndPuncttrue
	\mciteSetBstMidEndSepPunct{\mcitedefaultmidpunct}
	{\mcitedefaultendpunct}{\mcitedefaultseppunct}\relax
	\EndOfBibitem
	\bibitem[Zhao and Al{\`{u}}(2011)Zhao, and Al{\`{u}}]{Zhao2011}
	Zhao,~Y.; Al{\`{u}},~A. {Manipulating light polarization with ultrathin
		plasmonic metasurfaces}. \emph{Physical Review B} \textbf{2011}, \emph{84},
	205428\relax
	\mciteBstWouldAddEndPuncttrue
	\mciteSetBstMidEndSepPunct{\mcitedefaultmidpunct}
	{\mcitedefaultendpunct}{\mcitedefaultseppunct}\relax
	\EndOfBibitem
	\bibitem[{Balthasar Mueller} \latin{et~al.}(2017){Balthasar Mueller}, Rubin,
	Devlin, Groever, and Capasso]{BalthasarMueller2017}
	{Balthasar Mueller},~J.~P.; Rubin,~N.~A.; Devlin,~R.~C.; Groever,~B.;
	Capasso,~F. {Metasurface Polarization Optics: Independent Phase Control of
		Arbitrary Orthogonal States of Polarization}. \emph{Physical Review Letters}
	\textbf{2017}, \emph{118}, 113901\relax
	\mciteBstWouldAddEndPuncttrue
	\mciteSetBstMidEndSepPunct{\mcitedefaultmidpunct}
	{\mcitedefaultendpunct}{\mcitedefaultseppunct}\relax
	\EndOfBibitem
	\bibitem[Bosch \latin{et~al.}(2019)Bosch, Shcherbakov, Fan, and
	Shvets]{Bosch2019}
	Bosch,~M.; Shcherbakov,~M.~R.; Fan,~Z.; Shvets,~G. {Polarization states
		synthesizer based on a thermo-optic dielectric metasurface}. \emph{Journal of
		Applied Physics} \textbf{2019}, \emph{126}, 073102\relax
	\mciteBstWouldAddEndPuncttrue
	\mciteSetBstMidEndSepPunct{\mcitedefaultmidpunct}
	{\mcitedefaultendpunct}{\mcitedefaultseppunct}\relax
	\EndOfBibitem
	\bibitem[Ding \latin{et~al.}(2018)Ding, Pors, and Bozhevolnyi]{Ding2018}
	Ding,~F.; Pors,~A.; Bozhevolnyi,~S.~I. {Gradient metasurfaces: A review of
		fundamentals and applications}. \emph{Reports on Progress in Physics}
	\textbf{2018}, \emph{81}, 026401\relax
	\mciteBstWouldAddEndPuncttrue
	\mciteSetBstMidEndSepPunct{\mcitedefaultmidpunct}
	{\mcitedefaultendpunct}{\mcitedefaultseppunct}\relax
	\EndOfBibitem
	\bibitem[Fan \latin{et~al.}(2018)Fan, Shcherbakov, Allen, Allen, Wenner, and
	Shvets]{Fan2018b}
	Fan,~Z.; Shcherbakov,~M.~R.; Allen,~M.; Allen,~J.; Wenner,~B.; Shvets,~G.
	{Perfect Diffraction with Multiresonant Bianisotropic Metagratings}.
	\emph{ACS Photonics} \textbf{2018}, \emph{5}, 4303--4311\relax
	\mciteBstWouldAddEndPuncttrue
	\mciteSetBstMidEndSepPunct{\mcitedefaultmidpunct}
	{\mcitedefaultendpunct}{\mcitedefaultseppunct}\relax
	\EndOfBibitem
	\bibitem[Ni \latin{et~al.}(2013)Ni, Kildishev, and Shalaev]{Ni2013}
	Ni,~X.; Kildishev,~A.~V.; Shalaev,~V.~M. {Metasurface holograms for visible
		light}. \emph{Nature Communications} \textbf{2013}, \emph{4}, 2807\relax
	\mciteBstWouldAddEndPuncttrue
	\mciteSetBstMidEndSepPunct{\mcitedefaultmidpunct}
	{\mcitedefaultendpunct}{\mcitedefaultseppunct}\relax
	\EndOfBibitem
	\bibitem[Zheng \latin{et~al.}(2015)Zheng, M{\"{u}}hlenbernd, Kenney, Li,
	Zentgraf, and Zhang]{Zheng2015}
	Zheng,~G.; M{\"{u}}hlenbernd,~H.; Kenney,~M.; Li,~G.; Zentgraf,~T.; Zhang,~S.
	{Metasurface holograms reaching 80{\%} efficiency}. \emph{Nature
		Nanotechnology} \textbf{2015}, \emph{10}, 308--312\relax
	\mciteBstWouldAddEndPuncttrue
	\mciteSetBstMidEndSepPunct{\mcitedefaultmidpunct}
	{\mcitedefaultendpunct}{\mcitedefaultseppunct}\relax
	\EndOfBibitem
	\bibitem[Wang \latin{et~al.}(2016)Wang, Kruk, Tang, Li, Kravchenko, Neshev, and
	Kivshar]{Wang2016}
	Wang,~L.; Kruk,~S.; Tang,~H.; Li,~T.; Kravchenko,~I.; Neshev,~D.~N.;
	Kivshar,~Y.~S. {Grayscale transparent metasurface holograms}. \emph{Optica}
	\textbf{2016}, \emph{3}, 1504\relax
	\mciteBstWouldAddEndPuncttrue
	\mciteSetBstMidEndSepPunct{\mcitedefaultmidpunct}
	{\mcitedefaultendpunct}{\mcitedefaultseppunct}\relax
	\EndOfBibitem
	\bibitem[Khorasaninejad \latin{et~al.}(2016)Khorasaninejad, Zhu, Roques-Carmes,
	Chen, Oh, Mishra, Devlin, and Capasso]{Khorasaninejad2016}
	Khorasaninejad,~M.; Zhu,~A.~Y.; Roques-Carmes,~C.; Chen,~W.~T.; Oh,~J.;
	Mishra,~I.; Devlin,~R.~C.; Capasso,~F. {Polarization-Insensitive Metalenses
		at Visible Wavelengths}. \emph{Nano Letters} \textbf{2016}, \emph{16},
	7229--7234\relax
	\mciteBstWouldAddEndPuncttrue
	\mciteSetBstMidEndSepPunct{\mcitedefaultmidpunct}
	{\mcitedefaultendpunct}{\mcitedefaultseppunct}\relax
	\EndOfBibitem
	\bibitem[Khorasaninejad \latin{et~al.}(2016)Khorasaninejad, Chen, Devlin, Oh,
	Zhu, and Capasso]{Khorasaninejad2016a}
	Khorasaninejad,~M.; Chen,~W.~T.; Devlin,~R.~C.; Oh,~J.; Zhu,~A.~Y.; Capasso,~F.
	{Metalenses at visible wavelengths: Diffraction-limited focusing and
		subwavelength resolution imaging}. \emph{Science} \textbf{2016}, \emph{352},
	1190--1194\relax
	\mciteBstWouldAddEndPuncttrue
	\mciteSetBstMidEndSepPunct{\mcitedefaultmidpunct}
	{\mcitedefaultendpunct}{\mcitedefaultseppunct}\relax
	\EndOfBibitem
	\bibitem[Paniagua-Dom{\'{i}}nguez \latin{et~al.}(2018)Paniagua-Dom{\'{i}}nguez,
	Yu, Khaidarov, Choi, Leong, Bakker, Liang, Fu, Valuckas, Krivitsky, and
	Kuznetsov]{Paniagua-Dominguez2018}
	Paniagua-Dom{\'{i}}nguez,~R.; Yu,~Y.~F.; Khaidarov,~E.; Choi,~S.; Leong,~V.;
	Bakker,~R.~M.; Liang,~X.; Fu,~Y.~H.; Valuckas,~V.; Krivitsky,~L.~A.;
	Kuznetsov,~A.~I. {A Metalens with a Near-Unity Numerical Aperture}.
	\emph{Nano Letters} \textbf{2018}, \emph{18}, 2124--2132\relax
	\mciteBstWouldAddEndPuncttrue
	\mciteSetBstMidEndSepPunct{\mcitedefaultmidpunct}
	{\mcitedefaultendpunct}{\mcitedefaultseppunct}\relax
	\EndOfBibitem
	\bibitem[Khorasaninejad and Capasso(2017)Khorasaninejad, and
	Capasso]{Khorasaninejad2017}
	Khorasaninejad,~M.; Capasso,~F. {Metalenses: Versatile multifunctional photonic
		components}. \emph{Science} \textbf{2017}, \emph{358}, 1146\relax
	\mciteBstWouldAddEndPuncttrue
	\mciteSetBstMidEndSepPunct{\mcitedefaultmidpunct}
	{\mcitedefaultendpunct}{\mcitedefaultseppunct}\relax
	\EndOfBibitem
	\bibitem[Fu and Zhou(2010)Fu, and Zhou]{Fu2010}
	Fu,~Y.; Zhou,~X. {Plasmonic Lenses: A Review}. \emph{Plasmonics} \textbf{2010},
	\emph{5}, 287--310\relax
	\mciteBstWouldAddEndPuncttrue
	\mciteSetBstMidEndSepPunct{\mcitedefaultmidpunct}
	{\mcitedefaultendpunct}{\mcitedefaultseppunct}\relax
	\EndOfBibitem
	\bibitem[Wang \latin{et~al.}(2018)Wang, Wu, Su, Lai, Chen, Kuo, Chen, Chen,
	Huang, Wang, Lin, Kuan, Li, Wang, Zhu, and Tsai]{Wang2018}
	Wang,~S. \latin{et~al.}  {A broadband achromatic metalens in the visible}.
	\emph{Nature Nanotechnology} \textbf{2018}, \emph{13}, 227--232\relax
	\mciteBstWouldAddEndPuncttrue
	\mciteSetBstMidEndSepPunct{\mcitedefaultmidpunct}
	{\mcitedefaultendpunct}{\mcitedefaultseppunct}\relax
	\EndOfBibitem
	\bibitem[Shrestha \latin{et~al.}(2018)Shrestha, Overvig, Lu, Stein, and
	Yu]{Shrestha2018}
	Shrestha,~S.; Overvig,~A.~C.; Lu,~M.; Stein,~A.; Yu,~N. {Broadband achromatic
		dielectric metalenses}. \emph{Light: Science and Applications} \textbf{2018},
	\emph{7}, 85\relax
	\mciteBstWouldAddEndPuncttrue
	\mciteSetBstMidEndSepPunct{\mcitedefaultmidpunct}
	{\mcitedefaultendpunct}{\mcitedefaultseppunct}\relax
	\EndOfBibitem
	\bibitem[Kamali \latin{et~al.}(2016)Kamali, Arbabi, Arbabi, Horie, and
	Faraon]{Kamali2016a}
	Kamali,~S.~M.; Arbabi,~E.; Arbabi,~A.; Horie,~Y.; Faraon,~A. {Highly tunable
		elastic dielectric metasurface lenses}. \emph{Laser and Photonics Reviews}
	\textbf{2016}, \emph{10}, 1002--1008\relax
	\mciteBstWouldAddEndPuncttrue
	\mciteSetBstMidEndSepPunct{\mcitedefaultmidpunct}
	{\mcitedefaultendpunct}{\mcitedefaultseppunct}\relax
	\EndOfBibitem
	\bibitem[Zhou \latin{et~al.}(2020)Zhou, Shen, Li, Ge, Lu, and Hu]{Zhou2020}
	Zhou,~S.; Shen,~Z.; Li,~X.; Ge,~S.; Lu,~Y.; Hu,~W. {Liquid crystal integrated
		metalens with dynamic focusing property}. \emph{Optics Letters}
	\textbf{2020}, \emph{45}, 4324\relax
	\mciteBstWouldAddEndPuncttrue
	\mciteSetBstMidEndSepPunct{\mcitedefaultmidpunct}
	{\mcitedefaultendpunct}{\mcitedefaultseppunct}\relax
	\EndOfBibitem
	\bibitem[Afridi \latin{et~al.}(2018)Afridi, Canet-Ferrer, Philippet, Osmond,
	Berto, and Quidant]{Afridi2018}
	Afridi,~A.; Canet-Ferrer,~J.; Philippet,~L.; Osmond,~J.; Berto,~P.; Quidant,~R.
	{Electrically Driven Varifocal Silicon Metalens}. \emph{ACS Photonics}
	\textbf{2018}, \emph{5}, 4497--4503\relax
	\mciteBstWouldAddEndPuncttrue
	\mciteSetBstMidEndSepPunct{\mcitedefaultmidpunct}
	{\mcitedefaultendpunct}{\mcitedefaultseppunct}\relax
	\EndOfBibitem
	\bibitem[Chen \latin{et~al.}(2017)Chen, Feng, Monticone, Zhao, Zhu, Jiang,
	Zhang, Kim, Ding, Zhang, Al{\`{u}}, and Qiu]{Chen2017}
	Chen,~K.; Feng,~Y.; Monticone,~F.; Zhao,~J.; Zhu,~B.; Jiang,~T.; Zhang,~L.;
	Kim,~Y.; Ding,~X.; Zhang,~S.; Al{\`{u}},~A.; Qiu,~C.~W. {A Reconfigurable
		Active Huygens' Metalens}. \emph{Advanced Materials} \textbf{2017},
	\emph{29}, 1606422\relax
	\mciteBstWouldAddEndPuncttrue
	\mciteSetBstMidEndSepPunct{\mcitedefaultmidpunct}
	{\mcitedefaultendpunct}{\mcitedefaultseppunct}\relax
	\EndOfBibitem
	\bibitem[M{\"{u}}ller \latin{et~al.}(2002)M{\"{u}}ller, S{\"{o}}nnichsen, {Von
		Poschinger}, {Von Plessen}, Klar, and Feldmann]{Muller2002}
	M{\"{u}}ller,~J.; S{\"{o}}nnichsen,~C.; {Von Poschinger},~H.; {Von
		Plessen},~G.; Klar,~T.~A.; Feldmann,~J. {Electrically controlled light
		scattering with single metal nanoparticles}. \emph{Applied Physics Letters}
	\textbf{2002}, \emph{81}, 171--173\relax
	\mciteBstWouldAddEndPuncttrue
	\mciteSetBstMidEndSepPunct{\mcitedefaultmidpunct}
	{\mcitedefaultendpunct}{\mcitedefaultseppunct}\relax
	\EndOfBibitem
	\bibitem[Kossyrev \latin{et~al.}(2005)Kossyrev, Yin, Cloutier, Cardimona,
	Huang, Alsing, and Xu]{Kossyrev2005}
	Kossyrev,~P.~A.; Yin,~A.; Cloutier,~S.~G.; Cardimona,~D.~A.; Huang,~D.;
	Alsing,~P.~M.; Xu,~J.~M. {Electric field tuning of plasmonic response of
		nanodot array in liquid crystal matrix}. \emph{Nano Letters} \textbf{2005},
	\emph{5}, 1978--1981\relax
	\mciteBstWouldAddEndPuncttrue
	\mciteSetBstMidEndSepPunct{\mcitedefaultmidpunct}
	{\mcitedefaultendpunct}{\mcitedefaultseppunct}\relax
	\EndOfBibitem
	\bibitem[Gorkunov and Osipov(2008)Gorkunov, and Osipov]{Gorkunov2008}
	Gorkunov,~M.~V.; Osipov,~M.~A. {Tunability of wire-grid metamaterial immersed
		into nematic liquid crystal}. \emph{Journal of Applied Physics}
	\textbf{2008}, \emph{103}, 036101\relax
	\mciteBstWouldAddEndPuncttrue
	\mciteSetBstMidEndSepPunct{\mcitedefaultmidpunct}
	{\mcitedefaultendpunct}{\mcitedefaultseppunct}\relax
	\EndOfBibitem
	\bibitem[Xiao \latin{et~al.}(2009)Xiao, Chettiar, Kildishev, Drachev, Khoo, and
	Shalaev]{Xiao2009}
	Xiao,~S.; Chettiar,~U.~K.; Kildishev,~A.~V.; Drachev,~V.; Khoo,~I.~C.;
	Shalaev,~V.~M. {Tunable magnetic response of metamaterials}. \emph{Applied
		Physics Letters} \textbf{2009}, \emph{95}, 033115\relax
	\mciteBstWouldAddEndPuncttrue
	\mciteSetBstMidEndSepPunct{\mcitedefaultmidpunct}
	{\mcitedefaultendpunct}{\mcitedefaultseppunct}\relax
	\EndOfBibitem
	\bibitem[Kang \latin{et~al.}(2010)Kang, Woo, Choi, Lee, Kim, Kim, Hwang, Park,
	Kim, and Wu]{Kang2010}
	Kang,~B.; Woo,~J.~H.; Choi,~E.; Lee,~H.-H.; Kim,~E.~S.; Kim,~J.; Hwang,~T.-J.;
	Park,~Y.-S.; Kim,~D.~H.; Wu,~J.~W. {Optical switching of near infrared light
		transmission in metamaterial-liquid crystal cell structure}. \emph{Optics
		Express} \textbf{2010}, \emph{18}, 16492\relax
	\mciteBstWouldAddEndPuncttrue
	\mciteSetBstMidEndSepPunct{\mcitedefaultmidpunct}
	{\mcitedefaultendpunct}{\mcitedefaultseppunct}\relax
	\EndOfBibitem
	\bibitem[Decker \latin{et~al.}(2013)Decker, Kremers, Minovich, Miroshnichenko,
	Chigrin, Dragomir, Jagadish, and Kivshar]{Decker2013}
	Decker,~M.; Kremers,~C.; Minovich,~A.; Miroshnichenko,~A.~E.; Chigrin,~D.;
	Dragomir,~N.; Jagadish,~C.; Kivshar,~Y.~S. {Electro-optical switching by
		liquid-crystal controlled metasurfaces}. \emph{Optics Express} \textbf{2013},
	\emph{21}, 333--337\relax
	\mciteBstWouldAddEndPuncttrue
	\mciteSetBstMidEndSepPunct{\mcitedefaultmidpunct}
	{\mcitedefaultendpunct}{\mcitedefaultseppunct}\relax
	\EndOfBibitem
	\bibitem[Buchnev \latin{et~al.}(2013)Buchnev, Ou, Kaczmarek, Zheludev, and
	Fedotov]{Buchnev2013}
	Buchnev,~O.; Ou,~J.~Y.; Kaczmarek,~M.; Zheludev,~N.~I.; Fedotov,~V.~A.
	{Electro-optical control in a plasmonic metamaterial hybridised with a
		liquid-crystal cell}. \emph{Optics Express} \textbf{2013}, \emph{21},
	1633\relax
	\mciteBstWouldAddEndPuncttrue
	\mciteSetBstMidEndSepPunct{\mcitedefaultmidpunct}
	{\mcitedefaultendpunct}{\mcitedefaultseppunct}\relax
	\EndOfBibitem
	\bibitem[Atorf \latin{et~al.}(2014)Atorf, M{\"{u}}hlenbernd, Muldarisnur,
	Zentgraf, and Kitzerow]{Atorf2014}
	Atorf,~B.; M{\"{u}}hlenbernd,~H.; Muldarisnur,~M.; Zentgraf,~T.; Kitzerow,~H.
	{Electro-optic tuning of split ring resonators embedded in a liquid crystal}.
	\emph{Optics Letters} \textbf{2014}, \emph{39}, 1129\relax
	\mciteBstWouldAddEndPuncttrue
	\mciteSetBstMidEndSepPunct{\mcitedefaultmidpunct}
	{\mcitedefaultendpunct}{\mcitedefaultseppunct}\relax
	\EndOfBibitem
	\bibitem[Sautter \latin{et~al.}(2015)Sautter, Staude, Decker, Rusak, Neshev,
	Brener, and Kivshar]{Sautter2015}
	Sautter,~J.; Staude,~I.; Decker,~M.; Rusak,~E.; Neshev,~D.~N.; Brener,~I.;
	Kivshar,~Y.~S. {Active tuning of all-dielectric metasurfaces}. \emph{ACS
		Nano} \textbf{2015}, \emph{9}, 4308--4315\relax
	\mciteBstWouldAddEndPuncttrue
	\mciteSetBstMidEndSepPunct{\mcitedefaultmidpunct}
	{\mcitedefaultendpunct}{\mcitedefaultseppunct}\relax
	\EndOfBibitem
	\bibitem[Buchnev \latin{et~al.}(2015)Buchnev, Podoliak, Kaczmarek, Zheludev,
	and Fedotov]{Buchnev2015}
	Buchnev,~O.; Podoliak,~N.; Kaczmarek,~M.; Zheludev,~N.~I.; Fedotov,~V.~A.
	{Electrically Controlled Nanostructured Metasurface Loaded with Liquid
		Crystal: Toward Multifunctional Photonic Switch}. \emph{Advanced Optical
		Materials} \textbf{2015}, \emph{3}, 674--679\relax
	\mciteBstWouldAddEndPuncttrue
	\mciteSetBstMidEndSepPunct{\mcitedefaultmidpunct}
	{\mcitedefaultendpunct}{\mcitedefaultseppunct}\relax
	\EndOfBibitem
	\bibitem[Komar \latin{et~al.}(2017)Komar, Fang, Bohn, Sautter, Decker,
	Miroshnichenko, Pertsch, Brener, Kivshar, Staude, and Neshev]{Komar2017}
	Komar,~A.; Fang,~Z.; Bohn,~J.; Sautter,~J.; Decker,~M.; Miroshnichenko,~A.;
	Pertsch,~T.; Brener,~I.; Kivshar,~Y.~S.; Staude,~I.; Neshev,~D.~N.
	{Electrically tunable all-dielectric optical metasurfaces based on liquid
		crystals}. \emph{Applied Physics Letters} \textbf{2017}, \emph{110},
	071109\relax
	\mciteBstWouldAddEndPuncttrue
	\mciteSetBstMidEndSepPunct{\mcitedefaultmidpunct}
	{\mcitedefaultendpunct}{\mcitedefaultseppunct}\relax
	\EndOfBibitem
	\bibitem[Parry \latin{et~al.}(2017)Parry, Komar, Hopkins, Campione, Liu,
	Miroshnichenko, Nogan, Sinclair, Brener, and Neshev]{Parry2017}
	Parry,~M.; Komar,~A.; Hopkins,~B.; Campione,~S.; Liu,~S.;
	Miroshnichenko,~A.~E.; Nogan,~J.; Sinclair,~M.~B.; Brener,~I.; Neshev,~D.~N.
	{Active tuning of high-Q dielectric metasurfaces}. \emph{Applied Physics
		Letters} \textbf{2017}, \emph{111}, 053102\relax
	\mciteBstWouldAddEndPuncttrue
	\mciteSetBstMidEndSepPunct{\mcitedefaultmidpunct}
	{\mcitedefaultendpunct}{\mcitedefaultseppunct}\relax
	\EndOfBibitem
	\bibitem[Xie \latin{et~al.}(2017)Xie, Yang, Vashistha, Lee, and Chen]{Xie2017}
	Xie,~Z.-W.; Yang,~J.-H.; Vashistha,~V.; Lee,~W.; Chen,~K.-P. {Liquid-crystal
		tunable color filters based on aluminum metasurfaces}. \emph{Optics Express}
	\textbf{2017}, \emph{25}, 30764\relax
	\mciteBstWouldAddEndPuncttrue
	\mciteSetBstMidEndSepPunct{\mcitedefaultmidpunct}
	{\mcitedefaultendpunct}{\mcitedefaultseppunct}\relax
	\EndOfBibitem
	\bibitem[Shrekenhamer \latin{et~al.}(2013)Shrekenhamer, Chen, and
	Padilla]{Shrekenhamer2013}
	Shrekenhamer,~D.; Chen,~W.~C.; Padilla,~W.~J. {Liquid crystal tunable
		metamaterial absorber}. \emph{Physical Review Letters} \textbf{2013},
	\emph{110}, 177403\relax
	\mciteBstWouldAddEndPuncttrue
	\mciteSetBstMidEndSepPunct{\mcitedefaultmidpunct}
	{\mcitedefaultendpunct}{\mcitedefaultseppunct}\relax
	\EndOfBibitem
	\bibitem[Wang \latin{et~al.}(2017)Wang, Ge, Hu, Nakajima, and Lu]{Wang2017}
	Wang,~L.; Ge,~S.; Hu,~W.; Nakajima,~M.; Lu,~Y. {Graphene-assisted
		high-efficiency liquid crystal tunable terahertz metamaterial absorber}.
	\emph{Optics Express} \textbf{2017}, \emph{25}, 23873\relax
	\mciteBstWouldAddEndPuncttrue
	\mciteSetBstMidEndSepPunct{\mcitedefaultmidpunct}
	{\mcitedefaultendpunct}{\mcitedefaultseppunct}\relax
	\EndOfBibitem
	\bibitem[Komar \latin{et~al.}(2018)Komar, Paniagua-Dom{\'{i}}nguez,
	Miroshnichenko, Yu, Kivshar, Kuznetsov, and Neshev]{Komar2018}
	Komar,~A.; Paniagua-Dom{\'{i}}nguez,~R.; Miroshnichenko,~A.; Yu,~Y.~F.;
	Kivshar,~Y.~S.; Kuznetsov,~A.~I.; Neshev,~D. {Dynamic Beam Switching by
		Liquid Crystal Tunable Dielectric Metasurfaces}. \emph{ACS Photonics}
	\textbf{2018}, \emph{5}, 1742--1748\relax
	\mciteBstWouldAddEndPuncttrue
	\mciteSetBstMidEndSepPunct{\mcitedefaultmidpunct}
	{\mcitedefaultendpunct}{\mcitedefaultseppunct}\relax
	\EndOfBibitem
	\bibitem[Chung and Miller(2020)Chung, and Miller]{Chung2020}
	Chung,~H.; Miller,~O.~D. {Tunable Metasurface Inverse Design for 80{\%}
		Switching Efficiencies and 144° Angular Deflection}. \emph{ACS Photonics}
	\textbf{2020}, \emph{7}, 2236--2243\relax
	\mciteBstWouldAddEndPuncttrue
	\mciteSetBstMidEndSepPunct{\mcitedefaultmidpunct}
	{\mcitedefaultendpunct}{\mcitedefaultseppunct}\relax
	\EndOfBibitem
	\bibitem[Gorkunov \latin{et~al.}(2020)Gorkunov, Kasyanova, Artemov, Ezhov,
	Mamonova, Simdyankin, and Palto]{Gorkunov2020}
	Gorkunov,~M.~V.; Kasyanova,~I.~V.; Artemov,~V.~V.; Ezhov,~A.~A.;
	Mamonova,~A.~V.; Simdyankin,~I.~V.; Palto,~S.~P. {Superperiodic
		Liquid-Crystal Metasurfaces for Electrically Controlled Anomalous
		Refraction}. \emph{ACS Photonics} \textbf{2020},
	10.1021/acsphotonics.0c01168\relax
	\mciteBstWouldAddEndPuncttrue
	\mciteSetBstMidEndSepPunct{\mcitedefaultmidpunct}
	{\mcitedefaultendpunct}{\mcitedefaultseppunct}\relax
	\EndOfBibitem
	\bibitem[Shen \latin{et~al.}(2020)Shen, Zhou, Li, Ge, Chen, Hu, and
	Lu]{Shen2020}
	Shen,~Z.; Zhou,~S.; Li,~X.; Ge,~S.; Chen,~P.; Hu,~W.; Lu,~Y. {Liquid crystal
		integrated metalens with tunable chromatic aberration}. \emph{Advanced
		Photonics} \textbf{2020}, \emph{2}, 036002\relax
	\mciteBstWouldAddEndPuncttrue
	\mciteSetBstMidEndSepPunct{\mcitedefaultmidpunct}
	{\mcitedefaultendpunct}{\mcitedefaultseppunct}\relax
	\EndOfBibitem
	\bibitem[Li \latin{et~al.}(2019)Li, Xu, Veetil, Valuckas,
	Paniagua-Dom{\'{i}}nguez, and Kuznetsov]{Li2019}
	Li,~S.~Q.; Xu,~X.; Veetil,~R.~M.; Valuckas,~V.; Paniagua-Dom{\'{i}}nguez,~R.;
	Kuznetsov,~A.~I. {Phase-only transmissive spatial light modulator based on
		tunable dielectric metasurface}. \emph{Science} \textbf{2019}, \emph{364},
	1087--1090\relax
	\mciteBstWouldAddEndPuncttrue
	\mciteSetBstMidEndSepPunct{\mcitedefaultmidpunct}
	{\mcitedefaultendpunct}{\mcitedefaultseppunct}\relax
	\EndOfBibitem
	\bibitem[Wu \latin{et~al.}(2020)Wu, Shen, Ge, Chen, Shen, Wang, Zhang, Hu, Fan,
	Padilla, Lu, Jin, Chen, and Wu]{Wu2020}
	Wu,~J.; Shen,~Z.; Ge,~S.; Chen,~B.; Shen,~Z.; Wang,~T.; Zhang,~C.; Hu,~W.;
	Fan,~K.; Padilla,~W.; Lu,~Y.; Jin,~B.; Chen,~J.; Wu,~P. {Liquid crystal
		programmable metasurface for terahertz beam steering}. \emph{Applied Physics
		Letters} \textbf{2020}, \emph{116}, 131104\relax
	\mciteBstWouldAddEndPuncttrue
	\mciteSetBstMidEndSepPunct{\mcitedefaultmidpunct}
	{\mcitedefaultendpunct}{\mcitedefaultseppunct}\relax
	\EndOfBibitem
	\bibitem[Goodman(2005)]{Goodman2005}
	Goodman,~J.~W. \emph{{Introduction to Fourier optics}}; Roberts and Company
	Publishers, 2005\relax
	\mciteBstWouldAddEndPuncttrue
	\mciteSetBstMidEndSepPunct{\mcitedefaultmidpunct}
	{\mcitedefaultendpunct}{\mcitedefaultseppunct}\relax
	\EndOfBibitem
	\bibitem[Wu \latin{et~al.}(2014)Wu, Arju, Kelp, Fan, Dominguez, Gonzales,
	Tutuc, Brener, and Shvets]{wu_NatComm14}
	Wu,~C.; Arju,~N.; Kelp,~G.; Fan,~J.~A.; Dominguez,~J.; Gonzales,~E.; Tutuc,~E.;
	Brener,~I.; Shvets,~G. Spectrally selective chiral silicon metasurfaces based
	on infrared Fano resonances. \emph{Nat. Comm.} \textbf{2014}, \emph{5},
	3892\relax
	\mciteBstWouldAddEndPuncttrue
	\mciteSetBstMidEndSepPunct{\mcitedefaultmidpunct}
	{\mcitedefaultendpunct}{\mcitedefaultseppunct}\relax
	\EndOfBibitem
	\bibitem[Neuner~III \latin{et~al.}(2013)Neuner~III, Wu, Eyck, Sinclair, Brener,
	and Shvets]{Neuner2013}
	Neuner~III,~B.; Wu,~C.; Eyck,~G.~T.; Sinclair,~M.; Brener,~I.; Shvets,~G.
	Efficient infrared thermal emitters based on low-albedo polaritonic
	meta-surfaces. \emph{Applied Physics Letters} \textbf{2013}, \emph{102},
	211111\relax
	\mciteBstWouldAddEndPuncttrue
	\mciteSetBstMidEndSepPunct{\mcitedefaultmidpunct}
	{\mcitedefaultendpunct}{\mcitedefaultseppunct}\relax
	\EndOfBibitem
	\bibitem[Shcherbakov \latin{et~al.}(2019)Shcherbakov, Werner, Fan, Talisa,
	Chowdhury, and Shvets]{Shcherbakov2019}
	Shcherbakov,~M.~R.; Werner,~K.; Fan,~Z.; Talisa,~N.; Chowdhury,~E.; Shvets,~G.
	Photon acceleration and tunable broadband harmonics generation in nonlinear
	time-dependent metasurfaces. \emph{Nature communications} \textbf{2019},
	\emph{10}, 1--9\relax
	\mciteBstWouldAddEndPuncttrue
	\mciteSetBstMidEndSepPunct{\mcitedefaultmidpunct}
	{\mcitedefaultendpunct}{\mcitedefaultseppunct}\relax
	\EndOfBibitem
	\bibitem[Pryce \latin{et~al.}(2010)Pryce, Aydin, Kelaita, Briggs, and
	Atwater]{Pryce2010}
	Pryce,~I.~M.; Aydin,~K.; Kelaita,~Y.~A.; Briggs,~R.~M.; Atwater,~H.~A. {Highly
		strained compliant optical metamaterials with large frequency tunability}.
	\emph{Nano Letters} \textbf{2010}, \emph{10}, 4222--4227\relax
	\mciteBstWouldAddEndPuncttrue
	\mciteSetBstMidEndSepPunct{\mcitedefaultmidpunct}
	{\mcitedefaultendpunct}{\mcitedefaultseppunct}\relax
	\EndOfBibitem
	\bibitem[Yao \latin{et~al.}(2014)Yao, Shankar, Kats, Song, Kong, Loncar, and
	Capasso]{Yao2014}
	Yao,~Y.; Shankar,~R.; Kats,~M.~A.; Song,~Y.; Kong,~J.; Loncar,~M.; Capasso,~F.
	{Electrically tunable metasurface perfect absorbers for ultrathin
		mid-infrared optical modulators}. \emph{Nano Letters} \textbf{2014},
	\emph{14}, 6526--6532\relax
	\mciteBstWouldAddEndPuncttrue
	\mciteSetBstMidEndSepPunct{\mcitedefaultmidpunct}
	{\mcitedefaultendpunct}{\mcitedefaultseppunct}\relax
	\EndOfBibitem
	\bibitem[Wang \latin{et~al.}(2015)Wang, Zhang, Gu, Mehmood, Gong, Srivastava,
	Jian, Venkatesan, Qiu, and Hong]{Wang2015}
	Wang,~D.; Zhang,~L.; Gu,~Y.; Mehmood,~M.~Q.; Gong,~Y.; Srivastava,~A.;
	Jian,~L.; Venkatesan,~T.; Qiu,~C.~W.; Hong,~M. {Switchable Ultrathin
		Quarter-wave Plate in Terahertz Using Active Phase-change Metasurface}.
	\emph{Scientific Reports} \textbf{2015}, \emph{5}, 15020\relax
	\mciteBstWouldAddEndPuncttrue
	\mciteSetBstMidEndSepPunct{\mcitedefaultmidpunct}
	{\mcitedefaultendpunct}{\mcitedefaultseppunct}\relax
	\EndOfBibitem
	\bibitem[Ou \latin{et~al.}(2013)Ou, Plum, Zhang, and Zheludev]{Ou2013}
	Ou,~J.~Y.; Plum,~E.; Zhang,~J.; Zheludev,~N.~I. {An electromechanically
		reconfigurable plasmonic metamaterial operating in the near-infrared}.
	\emph{Nature Nanotechnology} \textbf{2013}, \emph{8}, 252--255\relax
	\mciteBstWouldAddEndPuncttrue
	\mciteSetBstMidEndSepPunct{\mcitedefaultmidpunct}
	{\mcitedefaultendpunct}{\mcitedefaultseppunct}\relax
	\EndOfBibitem
	\bibitem[Dabidian \latin{et~al.}(2015)Dabidian, Kholmanov, Khanikaev, Tatar,
	Trendafilov, Mousavi, Magnuson, Ruoff, and Shvets]{dabidian_ACSPhot14}
	Dabidian,~N.; Kholmanov,~I.; Khanikaev,~A.~B.; Tatar,~K.; Trendafilov,~S.;
	Mousavi,~S.~H.; Magnuson,~C.; Ruoff,~R.~S.; Shvets,~G. Electrical Switching
	of Infrared Light Using Graphene Integration with Plasmonic Fano Resonant
	Metasurfaces. \emph{ACS Photonics} \textbf{2015}, \emph{2}, 216--227\relax
	\mciteBstWouldAddEndPuncttrue
	\mciteSetBstMidEndSepPunct{\mcitedefaultmidpunct}
	{\mcitedefaultendpunct}{\mcitedefaultseppunct}\relax
	\EndOfBibitem
	\bibitem[Dabidian \latin{et~al.}(2016)Dabidian, Dutta-Gupta, Lai, Lu, Lee, Jin,
	Khanikaev, Kholmanov, , Mousavi, Magnuson, Ruoff, and
	Shvets]{dabidian_NanoLett16}
	Dabidian,~N.; Dutta-Gupta,~S.; Lai,~K.; Lu,~F.; Lee,~J.; Jin,~M.; Khanikaev,~S.
	T. A.~B.; Kholmanov,~I.; ; Mousavi,~S.~H.; Magnuson,~C.; Ruoff,~R.~S.;
	Shvets,~G. Experimental Demonstration of Phase Modulation and Motion Sensing
	Using Graphene-Integrated Metasurfaces. \emph{Nano Letters} \textbf{2016},
	\emph{16}, 3607--3615\relax
	\mciteBstWouldAddEndPuncttrue
	\mciteSetBstMidEndSepPunct{\mcitedefaultmidpunct}
	{\mcitedefaultendpunct}{\mcitedefaultseppunct}\relax
	\EndOfBibitem
	\bibitem[Love \latin{et~al.}(2009)Love, Hoffman, Hands, Gao, Kirby, and
	Banks]{Love2009}
	Love,~G.~D.; Hoffman,~D.~M.; Hands,~P.~J.; Gao,~J.; Kirby,~A.~K.; Banks,~M.~S.
	{High-speed switchable lens enables the development of a volumetric
		stereoscopic display}. \emph{Optics Express} \textbf{2009}, \emph{17},
	15716\relax
	\mciteBstWouldAddEndPuncttrue
	\mciteSetBstMidEndSepPunct{\mcitedefaultmidpunct}
	{\mcitedefaultendpunct}{\mcitedefaultseppunct}\relax
	\EndOfBibitem
	\bibitem[Li \latin{et~al.}(2006)Li, Mathine, Valley, {\"{A}}yr{\"{a}}s,
	Haddock, Giridhar, Williby, Schwiegerling, Meredith, Kippelen, Honkanen, and
	Peyghambarian]{Li2006}
	Li,~G.; Mathine,~D.~L.; Valley,~P.; {\"{A}}yr{\"{a}}s,~P.; Haddock,~J.~N.;
	Giridhar,~M.~S.; Williby,~G.; Schwiegerling,~J.; Meredith,~G.~R.;
	Kippelen,~B.; Honkanen,~S.; Peyghambarian,~N. {Switchable electro-optic
		diffractive lens with high efficiency for ophthalmic applications}.
	\emph{Proceedings of the National Academy of Sciences of the United States of
		America} \textbf{2006}, \emph{103}, 6100--6104\relax
	\mciteBstWouldAddEndPuncttrue
	\mciteSetBstMidEndSepPunct{\mcitedefaultmidpunct}
	{\mcitedefaultendpunct}{\mcitedefaultseppunct}\relax
	\EndOfBibitem
	\bibitem[Arbabi \latin{et~al.}(2015)Arbabi, Horie, Ball, Bagheri, and
	Faraon]{Arbabi2015}
	Arbabi,~A.; Horie,~Y.; Ball,~A.~J.; Bagheri,~M.; Faraon,~A.
	{Subwavelength-thick lenses with high numerical apertures and large
		efficiency based on high-contrast transmitarrays}. \emph{Nature
		Communications} \textbf{2015}, \emph{6}, 7069\relax
	\mciteBstWouldAddEndPuncttrue
	\mciteSetBstMidEndSepPunct{\mcitedefaultmidpunct}
	{\mcitedefaultendpunct}{\mcitedefaultseppunct}\relax
	\EndOfBibitem
	\bibitem[Cheng \latin{et~al.}(2006)Cheng, Chang, and Yeh]{Cheng2006}
	Cheng,~C.-C.; Chang,~C.~A.; Yeh,~J.~A. Variable focus dielectric liquid droplet
	lens. \emph{Optics Express} \textbf{2006}, \emph{14}, 4101--4106\relax
	\mciteBstWouldAddEndPuncttrue
	\mciteSetBstMidEndSepPunct{\mcitedefaultmidpunct}
	{\mcitedefaultendpunct}{\mcitedefaultseppunct}\relax
	\EndOfBibitem
	\bibitem[Tseng \latin{et~al.}(2018)Tseng, Hsiao, Chu, Chen, Sun, Liu, and
	Tsai]{Tseng2018}
	Tseng,~M.~L.; Hsiao,~H.-H.; Chu,~C.~H.; Chen,~M.~K.; Sun,~G.; Liu,~A.-Q.;
	Tsai,~D.~P. Metalenses: advances and applications. \emph{Advanced Optical
		Materials} \textbf{2018}, \emph{6}, 1800554\relax
	\mciteBstWouldAddEndPuncttrue
	\mciteSetBstMidEndSepPunct{\mcitedefaultmidpunct}
	{\mcitedefaultendpunct}{\mcitedefaultseppunct}\relax
	\EndOfBibitem
	\bibitem[Hong \latin{et~al.}(2017)Hong, Colburn, and Majumdar]{Hong2017}
	Hong,~C.; Colburn,~S.; Majumdar,~A. {Flat metaform near-eye visor}.
	\emph{Applied Optics} \textbf{2017}, \emph{56}, 8822--8827\relax
	\mciteBstWouldAddEndPuncttrue
	\mciteSetBstMidEndSepPunct{\mcitedefaultmidpunct}
	{\mcitedefaultendpunct}{\mcitedefaultseppunct}\relax
	\EndOfBibitem
	\bibitem[Sun \latin{et~al.}(2019)Sun, Xu, Sun, Valuckas, Zheng,
	Paniagua-Dom{\'\i}nguez, Kuznetsov]{Sun2019}
	Sun,~M.; Xu,~X.; Sun,~X.~W.; Liang,~X.; Valuckas, V.; Zheng,~Y.; Paniagua-Dom{\'\i}nguez,~R.; Kuznetsov,~A.~I. {  Efficient visible light modulation based on
	electrically tunable all dielectric metasurfaces embedded in thin-layer
	nematic liquid crystals}. \emph{Scientific Reports} \textbf{2019}, \emph{9},
	8673\relax
	\mciteBstWouldAddEndPuncttrue
	\mciteSetBstMidEndSepPunct{\mcitedefaultmidpunct}
	{\mcitedefaultendpunct}{\mcitedefaultseppunct}\relax
	\EndOfBibitem
	\bibitem[Cambiasso \latin{et~al.}(2017)Cambiasso, Grinblat, Li, Rakovich,
	Cort{\'{e}}s, and Maier]{Cambiasso2017}
	Cambiasso,~J.; Grinblat,~G.; Li,~Y.; Rakovich,~A.; Cort{\'{e}}s,~E.;
	Maier,~S.~A. {Bridging the Gap between Dielectric Nanophotonics and the
		Visible Regime with Effectively Lossless Gallium Phosphide Antennas}.
	\emph{Nano Letters} \textbf{2017}, \emph{17}, 1219--1225\relax
	\mciteBstWouldAddEndPuncttrue
	\mciteSetBstMidEndSepPunct{\mcitedefaultmidpunct}
	{\mcitedefaultendpunct}{\mcitedefaultseppunct}\relax
	\EndOfBibitem
	\bibitem[Devlin \latin{et~al.}(2016)Devlin, Khorasaninejad, Chen, Oh, and
	Capasso]{Devlin2016}
	Devlin,~R.~C.; Khorasaninejad,~M.; Chen,~W.~T.; Oh,~J.; Capasso,~F. {Broadband
		high-efficiency dielectric metasurfaces for the visible spectrum}.
	\emph{Proceedings of the National Academy of Sciences of the United States of
		America} \textbf{2016}, \emph{113}, 10473--10478\relax
	\mciteBstWouldAddEndPuncttrue
	\mciteSetBstMidEndSepPunct{\mcitedefaultmidpunct}
	{\mcitedefaultendpunct}{\mcitedefaultseppunct}\relax
	\EndOfBibitem
	\bibitem[Miller \latin{et~al.}(2017)Miller, Yu, Ji, Griffith, Cardenas, Gaeta,
	and Lipson]{Miller2017}
	Miller,~S.~A.; Yu,~M.; Ji,~X.; Griffith,~A.~G.; Cardenas,~J.; Gaeta,~A.~L.;
	Lipson,~M. {Low-loss silicon platform for broadband mid-infrared photonics}.
	\emph{Optica} \textbf{2017}, \emph{4}, 707\relax
	\mciteBstWouldAddEndPuncttrue
	\mciteSetBstMidEndSepPunct{\mcitedefaultmidpunct}
	{\mcitedefaultendpunct}{\mcitedefaultseppunct}\relax
	\EndOfBibitem
	\bibitem[Yu \latin{et~al.}(2015)Yu, Zhu, Paniagua-Dom{\'{i}}nguez, Fu,
	Luk'yanchuk, and Kuznetsov]{Yu2015}
	Yu,~Y.~F.; Zhu,~A.~Y.; Paniagua-Dom{\'{i}}nguez,~R.; Fu,~Y.~H.;
	Luk'yanchuk,~B.; Kuznetsov,~A.~I. {High-transmission dielectric metasurface
		with 2$\pi$ phase control at visible wavelengths}. \emph{Laser {\&} Photonics
		Reviews} \textbf{2015}, \emph{9}, 412--418\relax
	\mciteBstWouldAddEndPuncttrue
	\mciteSetBstMidEndSepPunct{\mcitedefaultmidpunct}
	{\mcitedefaultendpunct}{\mcitedefaultseppunct}\relax
	\EndOfBibitem
	\bibitem[Hu \latin{et~al.}(2019)Hu, Luo, Chen, Liu, Li, Wang, Liu, and
	Duan]{Hu2019}
	Hu,~Y.; Luo,~X.; Chen,~Y.; Liu,~Q.; Li,~X.; Wang,~Y.; Liu,~N.; Duan,~H.
	{3D-Integrated metasurfaces for full-colour holography}. \emph{Light: Science
		and Applications} \textbf{2019}, \emph{8}, 86\relax
	\mciteBstWouldAddEndPuncttrue
	\mciteSetBstMidEndSepPunct{\mcitedefaultmidpunct}
	{\mcitedefaultendpunct}{\mcitedefaultseppunct}\relax
	\EndOfBibitem
\end{mcitethebibliography}
\end{document}